\documentclass[prl,twocolumn,,showpacs,superscriptaddress,floatfix, nofootinbib]{revtex4-2}

\usepackage{amsmath,amssymb,amsfonts,mathrsfs, amsbsy}
\usepackage[utf8]{inputenc}
\usepackage[T1]{fontenc}

\usepackage{xcolor} % in preamble
\usepackage{graphicx}
\usepackage[bookmarks=true, colorlinks=true, linkcolor=blue, urlcolor=blue, citecolor=blue, bookmarks=true, hyperindex=true]{hyperref}
\usepackage[normalem]{ulem}
\usepackage{comment}
\usepackage{array}
\usepackage{float}  
\usepackage{makecell}
\usepackage{enumitem}

\usepackage{bbm}

\newcommand{\be}{\begin{equation}}
\newcommand{\ee}{\end{equation}}
\newcommand{\beq}{\begin{eqnarray}}
\newcommand{\eeq}{\end{eqnarray}}

\begin{document}

\title{Effective field theories of nonlinear fluctuating hydrodynamics in one dimension}

\author{Matija Koterle}

\author{Enej Ilievski}
\thanks{Corresponding author: enej.ilievski@fmf.uni-lj.si}
\affiliation{Faculty for Mathematics and Physics, University of Ljubljana, Jadranska 19, SI-1000 Ljubljana, Slovenia}

%\author{Matija Koterle and Enej Ilievski\\[1ex]
%\small Faculty for Mathematics and Physics, University of Ljubljana, Jadranska 19, SI-1000 Ljubljana, Slovenia}

\begin{abstract}
Describing emergent macroscopic phenomena in low spatial dimensions is known to be notoriously challenging, primarily due to strong interactions that render perturbative approaches inapplicable. On the other hand, low-dimensional systems host a wealth of unorthodox phenomena. A prominent example is the emergence of superdiffusive transport in one-dimensional interacting systems, traditionally studied in the framework of nonlinear fluctuating hydrodynamics. After identifying and discussing internal inconsistencies in the previous formulations, in this work we develop a general and systematic approach for constructing effective field theories of one-dimensional hydrodynamic systems in the form of coupled stochastic Langevin-type equations compatible with the physical requirements of local equilibrium states --- the thermodynamic Maxwell relation and fluctuation-dissipation symmetry. To exemplify our construction, we devise a general numerical integration scheme and implement it on a simple model of two interacting hydrodynamic modes with a non-Gaussian stationary equilibrium measure.
\end{abstract}

\maketitle

\paragraph*{\textbf{Introduction.}} A rich variety of problems in statistical mechanics concerning the large-scale behavior of complex interacting systems can be formulated in terms of stochastic field theories. These provide an effective description of relevant collective degrees of freedom, and have proven instrumental in quantitative studies of macroscopic emergent phenomena, including critical behavior near phase transitions \cite{Hohenberg_1977}, the glass transition \cite{Berthier_2011}, kinetic roughening \cite{Halpin_Healy_1995,Krug_1997}, and the transport behavior in driven diffusive systems \cite{MFT}.

One of the most rudimentary applications of statistical field theory is hydrodynamics, an effective theory for a collection of slowly relaxing, coarse-grained, macroscopic densities $\rho(x,t)$ governed by a continuity equation $\partial_{t}\rho(x,t)+\nabla \cdot \vec{\mathcal{J}}[\rho(x,t)]=0$. The hydrodynamic current $\vec{\mathcal{J}}[\rho(x,t)]$ generally includes dissipative terms resulting from microscopic degrees of freedom that have been discarded and enter macroscopically as a stochastic force.

In one spatial dimension, single-species particle-conserving systems feature two distinct dynamical behaviors: the Edwards--Wilkinson universality class of the stochastic diffusion equation and the Kardar--Parisi--Zhang (KPZ) \cite{Kardar_1986} universality of the stochastic Burgers equation, governing dynamic roughening of growing interfaces \cite{Quastel_2015,Corwin_2016,Takeuchi_2018} and charge-conserving quantum circuits \cite{Sharipov_2026}.
Systems with multiple hydrodynamic modes can however exhibit more intricate dynamical properties. An emblematic example is the chain of oscillators interacting via an anharmonic potential, known as the Fermi--Pasta--Ulam--Tsingou (FPUT) model, which features sound and heat modes that undergo superdiffusive spreading, see Refs.~\cite{Lepri_1997,Lepri_1998,Lepri_2003,Spohn_2006,Mai_2007,spohn2014nonlinear,lepri2026anomalous,minami2026symmetry}.

%Their \emph{equilibrium} dynamical structure factors $S_{a}(x,t)$ reveal \emph{anomalous} (i.e. non-Gaussian) broadening (in the comoving frame) at late times, $\int dx x^{2} S_{a}(x-v_{a}t,t)\sim t^{2/z_{a}}$, characterized by their own superdiffusive algebraic exponents $z_{a}<2$.
To explain the emergence of superdiffusive broadening of sound and heat peaks in the FPUT chain, Ref.\,\cite{spohn2014nonlinear} introduced an effective field theory of nonlinear fluctuating hydrodynamics (NLFHD). Written in terms of hydrodynamic (normal) modes $\boldsymbol{\phi}\equiv \{\phi_{a}(x,t)\}$, representing small fluctuations above an equilibrium background, NLFHD constitutes a system of coupled \emph{nonlinear} stochastic Langevin equations,
\begin{equation} 
    \partial_{t}\phi_{a}(x,t) + \partial_{x}\big(\mathcal{J}^{\rm R}_{a}[\boldsymbol{\phi}]+\mathcal{J}^{\rm D}_{a}[\boldsymbol{\phi}]+\xi_{a}(x,t)\big) = 0,
    \label{eq:NLFHD}
\end{equation}
where 
\begin{equation}
    \label{eq:NLFHD_currents}
    \mathcal{J}^{\rm R}_{a} \equiv \mathrm{v}_{a}\phi_{a} + \sum_{b,c}\mathrm{G}^{a}_{bc}\phi_{b}\phi_{c},\quad
    \mathcal{J}^{\rm D}_{a} \equiv -\sum_{b}\mathrm{D}_{ab}\partial_{x}\phi_{b},
\end{equation}
and $\xi_{a}(x,t)$ represent the additive white Gaussian noise of zero mean and covariance $\sigma_{ab}=2\mathrm{D}_{ab}$ set by the fluctuation-dissipation relation (FDR). Crucially, the reversible (i.e. conservative) component $\mathcal{J}^{\rm R}_{a}$ includes additional interacting terms supplied by the mode-coupling tensor $\mathrm{G}$ encoding the Hessian of equilibrium currents.

In this Letter, we revisit the foundational principles of nonlinear fluctuating hydrodynamics and highlight several critical conceptual issues. Our main contribution is a general theoretical framework for constructing consistent hydrodynamic effective field theories compatible with the fundamental symmetry constraints of local equilibrium. We specialize in one spatial dimension where nonlinearities play a pivotal role. Moreover, we devise and implement a numerical integration scheme and benchmark it on a minimal model with a non-Gaussian equilibrium measure. Our construction enables quantitative investigation of transport properties beyond the previous approaches.

\paragraph*{\textbf{Issues with nonlinear fluctuating hydrodynamics.}}

Before outlining our construction, it is essential to carefully examine the setting and central assumption of conventional NLFHD.
The standard approach \cite{mendl2013,spohn2014nonlinear,spohn2016fluctuating} based on Eqs.~\eqref{eq:NLFHD} and \eqref{eq:NLFHD_currents} has recently been subjected to scrutiny in Ref.\,\cite{minami2026symmetry}, which designed an alternative, symmetry-based, reformulation of NLFHD: by postulating a Gaussian stationary equilibrium measure, $\mathcal{P}_{\rm eq}\simeq \exp{(-\tfrac{1}{2}\sum_a\int dx\,\phi^{2}_{a})}$, the bare diffusion matrix $\mathrm{D}$ and couplings $\mathrm{G}$ in Eqs.~\eqref{eq:NLFHD_currents} are calibrated to ensure the FDR and stationarity, respectively.
Based on the dynamical renormalization group analysis, \,\cite{minami2026symmetry} concludes that the stationary (heat) mode exhibits KPZ scaling, in tension with both previous theoretical predictions \cite{mendl2013,spohn2014nonlinear,spohn2016fluctuating} and microscopic simulations \cite{mendl2013}.

We argue now that neither the standard nor the symmetry-based approach of \cite{minami2026symmetry} are adequate to quantitatively describe transport properties in generic one-dimensional systems featuring interacting sound modes. The crux of the matter is that
\begin{enumerate}[label=\roman*]
    \item \emph{the reversible component $\mathcal{J}^{\rm R}_{a}[\boldsymbol{\phi}]$ of the form \eqref{eq:NLFHD_currents} with generic $\mathrm{G}$ does not preserve the Gaussian measure}, and
    \item \emph{the Gaussian measure $\mathcal{P}_{\rm eq}$ in generically incompatible with static cumulants in equilibrium states of the underlying microscopic model.}
\end{enumerate}
Indeed, the higher-order nonlinearities in $\mathcal{J}^{\rm R}_{a}[\boldsymbol{\phi}]$, which are commonly discarded as irrelevant (in the sense of renormalization group flow \cite{minami2026symmetry}) terms, are generically present and thus induce a drift in the initial Gaussian measure; including such terms is essential for maintaining detailed balance in the hydrodynamic EFT at all times. In addition to that, the coupling coefficients entering both the reversible and dissipative components of the hydrodynamic currents represent bare quantities (resulting from the eliminated non-hydrodynamic modes), and must thus be distinguished from \emph{dressed} quantities that represent true, physical transport coefficients of the microscopic theory. For instance, velocities $\mathrm{v}_{a}$ appearing in Eq.~\eqref{eq:NLFHD_currents} do not coincide with the sound-peak velocities $v_{a}$ measured by dynamical structure factors. Dressing in fact affects generic quantities and occurs even for Gaussian stationary measures. 
Identifying $\mathrm{G}$ with the equilibrium current Hessian (rotated to $\boldsymbol{\phi}$-basis) of the microscopic model, as assumed in Refs.\cite{spohn2014nonlinear,minami2026symmetry}, indeed requires $\mathcal{P}_{\rm eq}$ to be Gaussian; consequently, stationarity of $\mathcal{P}_{\rm eq}$ under $\mathcal{J}_{a}[\boldsymbol{\phi}]$ requires $\mathrm{G}$ to be a \emph{completely} symmetric tensor. This property is, however, generally incompatible with the structure of microscopic currents, including the FPUT chains.

The other ingredient of NLFHD is the (Navier--Stokes) dissipative component $\mathcal{J}^{\rm D}_{a}[\boldsymbol{\phi}]$, cf. Eq.~\eqref{eq:NLFHD_currents} that is introduced on the phenomenological ground. To obtain a correct quantitative prediction of asymptotically Gaussian structure factors (or sub-leading corrections to superdiffusive modes) the so-called bare diffusion coefficients $\mathrm{D}$ must be determined from the dynamical correlations of microscopic current densities using, for example, a projected Kubo formula \cite{Saito_2021,doyon2025hydrodynamic}.

To our knowledge, one-dimensional NLFHD (cf. Eqs.~\eqref{eq:NLFHD}) have predominantly been numerically solved in multi-component hydrodynamic systems with a Gaussian equilibrium measure with a specific measure-preserving discretization \cite{PhysRevLett.131.197102,roy2024universality, Schmidt_2025, Schmidt_2026, minami2026symmetry}. Indeed, as shown in Supplemental Material \cite{SM}, numerical integration of Eqs.~\eqref{eq:NLFHD} initialized in a Gaussian state using such an integrator can result in a spurious, uncontrolled, dynamical drift to a non-Gaussian state or even lead to a blow-up.
In most applications, Eqs.~\eqref{eq:NLFHD} are indeed bypassed, replaced by a system of \emph{deterministic} PDEs for dynamical two-time functions (structure factors) known as the \emph{mode-coupling equations} (MCE) \cite{van_Beijeren_2012}. As detailed out in End Matter, the standard MCE likewise rest on certain assumptions that are not fulfilled in generic hydrodynamic EFTs.

The remaining discussion is organized as follows. After reviewing the key symmetry relations satisfied by local equilibrium states, we outline how to construct consistent hydrodynamic EFTs built on the same symmetry principles. Finally, we present an explicit numerical implementation of our approach for a particular class of two-component systems that features superdiffusive transport.

\paragraph*{\textbf{Thermodynamics and Maxwell relation.}}
We consider one-dimensional classical systems with local interactions that possess a finite number of local conservation laws $\mathrm{Q}_{a}=\int dx\,\mathrm{q}_{a}(x)$. 
The primary objective is to obtain an effective equation of motion for small perturbations near local equilibrium states $\rho_{\rm eq} \simeq \exp{(-\sum_{i}\mu_{i}\mathrm{Q}_{i})}$ with chemical potentials $\boldsymbol{\mu}=\{\mu_{i}\}$. The local continuity equation, $\partial_{t}\mathrm{q}_{a}(x,t)+\partial_{x}\mathrm{j}_{a}(x,t)=0$, implies that charge density fluctuations undergo (algebraically) slow relaxation in time. Let $\boldsymbol{q}\equiv \{q_{a}\}$, with $q_{a}(\boldsymbol{\mu})=\langle \mathrm{q}_{a}\rangle_{\boldsymbol{\mu}}$, designate average equilibrium charge densities. By fixing the gauge, we set $j_{a}(\boldsymbol{\mu})=\langle \mathrm{j}_{a}\rangle_{\boldsymbol{\mu}}=0$. In terms of thermodynamic potentials, the free-energy $f(\boldsymbol{\mu})$ and `free-energy flux' $g(\boldsymbol{\mu})$, the charge and current averages are given by $q_{a}=-\partial f(\boldsymbol{\mu})/\partial \mu_{a}$ and $j_{a}=-\partial g(\boldsymbol{\mu})/\partial \mu_{a}$, respectively, with the corresponding equal-time (i.e. static) covariance matrices $C(\boldsymbol{\mu})$ and $B(\boldsymbol{\mu})$ reading $C_{ab} \equiv \int dx\langle \mathrm{q}_{a}(x)\mathrm{q}_{b}(0) \rangle^{c}_{\boldsymbol{\mu}}=-\partial_{\mu_{b}}q_{a}(\boldsymbol{\mu})$ and $B_{a|b} \equiv \int dx\langle \mathrm{j}_{a}(x)\mathrm{q}_{b}(0) \rangle^{c}_{\boldsymbol{\mu}}=-\partial_{\mu_{b}}j_{a}(\boldsymbol{\mu})$ (here and subsequently we use a bar to split the index referring to the current from the charge, permutationally invariant indices). While the symmetry $C=C^{T}$ is manifest, the \emph{Maxwell symmetry}
\begin{equation}
    \label{eq:Maxwell_symmetry}
    \frac{\partial j_{a}(\boldsymbol{\mu})}{\partial \mu_{b}}=\frac{\partial j_{b}(\boldsymbol{\mu})}{\partial \mu_{a}}\qquad \Rightarrow\qquad B(\boldsymbol{\mu})=B(\boldsymbol{\mu})^{T},    
\end{equation}
is a corollary of space-time stationarity which implies existence of $g(\boldsymbol{\mu})$ \cite{Grisi_2011,Doyon_2017,Karevski_2019,Durnin_2021}. 

%The effective velocities $v_{a}$ of propagating linearized fluctuations above a uniform equilibrium background are obtained by rotating $\mathrm{q}_{b}$ to new densities $\mathrm{n}_{a}(x,t)$ via  $\mathrm{n}_{a}=\sum_{b}R_{ab}\mathrm{q}_{b}$, where $R$ transforms the current Jacobian $A=BC^{-1}$, with $A_{a|b}=[\partial j_{a}(\boldsymbol{\mu})/\partial q_{b}(\boldsymbol{\mu})]_{\rho_{\rm eq}}$, into diagonal form, i.e. $A=R^{-1}VR$ with $V={\rm diag}(v_{a})$.
%We choose the standard normalization $RCR^{T}=1$ such that $\mathcal{C}_{ab}\equiv \int dx\langle \mathrm{n}_{a}(x,t)\mathrm{n}_{b}(0,t)\rangle^{c}_{\rho_{\rm eq}}=\delta_{ab}$, and use calligraphic symbols to write tensors in the normal-mode basis. Analogously, one can define the higher-point static correlators $\mathcal{B}_{a|b_{1}\ldots b_{n}}$ and $\mathcal{C}_{a_{1}\ldots a_{n}}$.

Since static covariances are analytic functions of equilibrium charge densities $\mathbf{q}=\{q_{a}\}$ dual to $\boldsymbol{\mu}$, the Maxwell symmetry, $\varDelta_{B}(\mathbf{q})\equiv B^{T}(\mathbf{q})-B(\mathbf{q})=0$, generates an infinite tower of relations through iterative differentiation, $\partial_{\mathbf{q}}^{n}\varDelta_{B}(\mathbf{q})|_{\boldsymbol{\mu}} = 0$.

\paragraph*{\textbf{Hydrodynamic EFTs.}}
Hydrodynamic EFTs govern the large-scale dynamics of charge-density fluctuations $\psi_{a}(x,t)$ above a homogeneous equilibrium state with average densities $\mathbf{q}$. Constructing an
EFT relies on a particular scale separation, $\ell_{\rm micro}\ll \ell_{\rm meso}\ll \lambda$: characteristic wavelengths $\lambda$ of hydrodynamic modes $\psi_{a}(x,t)$ must be large compared to a mesoscopic `fluid cell' of size $\ell_{\rm meso}$ at \emph{macroscopic} coordinate $x$; fluid cells are large compared to microscopic scales $\ell_{\rm micro}$ and assumed to be in local equilibrium.
Hydrodynamic fields $\psi_{a}(x,t)$ thus represent coarse-grained charge densities beyond $\ell_{\rm meso}$ (i.e. in the absence of non-hydrodynamic modes) subject to an effective \emph{stochastic} evolution law.

Explicit elimination of non-hydrodynamic modes is unfortunately not feasible in practice, and therefore building hydrodynamic EFTs requires a `top-down approach': hydrodynamic currents are represented as a formally infinite double series, organized with respect to orders of gradients and nonlinearities, compatible with the symmetry constraints of the microscopic theory (see Refs.~\cite{Crossley_2017,glorioso2018lectures,Glorioso_2021,Michailidis_2024} for a modern Schwinger--Keldysh formulation).

While similar in spirit, here we outline a different approach rooted in the Maxwell relation \eqref{eq:Maxwell_symmetry} that restricts the conservative sector of the theory.
The connection with a target microscopic model is established based on the `UV-IR matching', i.e. by requiring that at large times and distances the dynamical correlation functions of microscopic densities match the hydrodynamic correlators:
\begin{equation}
    \big\langle \prod_{i=1}^{n}\mathrm{q}_{a_{i}}(x_{a_{i}},t_{a_{i}}) \big\rangle^{c}_{\rho_{\rm eq}}
    \asymp \big\langle \prod_{i=1}^{n}\psi_{a_{i}}(x_{a_{i}},t_{a_{i}}) \big\rangle^{c}_{\mathcal{P}_{\rm eq},\boldsymbol{\xi}}.
\end{equation}
Crucially, statistical averaging in the EFT is over the stationary equilibrium measure $\mathcal{P}_{\rm eq}[\boldsymbol{\psi}]$ and, additionally, over stochastic (i.e. noise) fields $\boldsymbol{\xi}$, using the prescription $\langle \mathcal{O}\rangle_{\boldsymbol{\xi}}=\int \mathscr{D}\boldsymbol{\xi}\,\mathcal{P}_{\xi}[\boldsymbol{\xi}]\, \mathcal{O}$, with a Gaussian measure $\mathcal{P}_{\xi}[\boldsymbol{\xi}]\simeq \exp{(-\tfrac{1}{2}\sum_{a,b}\int \mathrm{d}t\,dx\,\xi_{a}(x,t)[(\sigma^{\flat})^{-1}]_{ab}\xi_{b}(x,t))}$. Dynamical expectation values are then computed as double averages over initial field configurations $\boldsymbol{\psi}_0$ and noise realizations $\boldsymbol{\xi}$,
\begin{equation}
    \langle \mathcal{O}(t) \rangle_{\mathcal{P}_{0},\boldsymbol{\xi}}\equiv \int \mathscr{D}\boldsymbol{\psi}_{0}\,\mathcal{P}_{0}[\boldsymbol{\psi}_{0}]\langle \mathcal{O}[\boldsymbol{\psi}_t]\rangle_{\boldsymbol{\xi}}, 
\end{equation}
The average of the observable over noise is $\langle \mathcal{O}[\boldsymbol{\psi}_t]\rangle_{\boldsymbol{\xi}}\equiv \int \mathcal{D}\boldsymbol{\xi} \, \mathcal{P}_{\xi}[\boldsymbol{\xi}] \,\mathcal{O}[\psi_t]$, where $\mathcal{O}[\psi_t]$ is a short-hand notation, denoting a time-evolved state $\boldsymbol{\psi}_t=\boldsymbol{\psi}_t[\boldsymbol{\psi}_0;\boldsymbol{\xi}]$ conditioned on the initial configuration and noise realization.
\begin{comment}
\begin{equation}
    \langle \mathcal{O}(t) \rangle_{\mathcal{P}_{0},\boldsymbol{\xi}}\equiv \int \mathscr{D}\boldsymbol{\psi}_{0}\,\mathcal{P}_{0}[\boldsymbol{\psi}_{0}]\langle \mathcal{O}_t[\boldsymbol{\psi}_0]\rangle_{\boldsymbol{\xi}}, 
\end{equation}
where $\mathcal{O}_t[\boldsymbol{\psi_0}]\equiv \mathcal{O}[\boldsymbol{\psi}_t]$ denotes an observable evaluated in a time-evolved state $\boldsymbol{\psi}_t$ with an initial configuration $\boldsymbol{\psi}_0$.
\end{comment}

%$\langle \mathcal{O}[\boldsymbol{\psi}] \rangle_{\mathcal{P}_{\rm eq}}\equiv \int \mathscr{D}\boldsymbol{\psi}\,\mathcal{P}_{\rm eq}[\boldsymbol{\psi}]\,\mathcal{O}[\boldsymbol{\psi}]$ denotes the functional average with respect to the stationary equilibrium measure $\mathcal{P}_{\rm eq}$ of the EFT, which plays a pivotal role in our formalism.

We proceed by outlining a general formulation of hydrodynamic EFTs. Our chief goal here is to develop a formalism for constructing consistent EFTs, and not to describe any particular microscopic model. To this end, we can afford to make certain technical simplifications. In particular, we discard higher-order gradient terms which are inessential for capturing relaxation to the leading asymptotic order.
%Our aim here is to engineer EFTs that are specifically tailored for capturing the late-time relaxation of normal modes to leading order in an asymptotic series (we discard the gradient terms in $\mathcal{J}^{\rm R}_{a}[\boldsymbol{\psi}]$ and similarly the higher-order (Burnett-type) terms entering $\mathcal{J}^{\rm D}_{a}[\boldsymbol{\psi}]$, which can be crucial for properly capturing relaxation dynamics at smaller wavelengths).
Unlike conventional NLFHD, it is, however, pivotal to accommodate non-Gaussian measures and higher-order nonlinearities in $\mathcal{J}^{\rm R}_{a}[\boldsymbol{\psi}]$.
%For definiteness, we hereafter specialize to a particular class of EFTs designed to capture broadening of non-degenerate sound modes to the leading order in asymptotic series. This requires including quadratic nonlinearities in the reversible current $\mathcal{J}^{\rm R}_{a}[\boldsymbol{\psi}]$ and dissipation to leading linear order in gradients.

The first step is to write the reversible component of the hydrodynamic current in the form
\begin{equation}
    \mathcal{J}^{\rm R}_{a}[\boldsymbol{\psi}] =
    \sum_{n \geq1}
    \frac{1}{n!}
    \sum_{b_{1},\ldots,b_{n}}
    A^{\flat}_{a|b_{1},\ldots,b_{n}}
    \prod_{i=1}^{n}\psi_{b_{i}}(x),
\end{equation}
\noindent
where $A_{a|b_1,\ldots, b_n}^\flat$ are tensors of rank $n+1$, symmetric in $b$-indices.
Here and subsequently we use the subscript $\flat$ to distinguish \emph{bare} couplings from physical, dressed ones.
Similarly, we represent the normalized stationary measure $\mathcal{P}_{\rm eq}[\boldsymbol{\psi}]\simeq e^{-\mathcal{S}_{\rm eq}[\boldsymbol{\psi}]}$ by expanding the `entropy' functional $\mathcal{S}_{\rm eq}=\int dx \mathscr{S}[\boldsymbol{\psi}]$,
\begin{equation}
    \label{eq:equilibrium_entropy}
    \mathscr{S}[\boldsymbol{\psi}] = \sum_{n\geq 1}\frac{1}{n!}\sum_{a_{1},\ldots,a_{n}}K^{\flat}_{a_{1},\ldots,a_{n}}\prod_{i=1}^{n}\psi_{a_{i}}(x),
\end{equation}
where $K^\flat_{a_{1},\ldots,a_{n}}$ are completely symmetric tensors of rank $n$
which, importantly, do \emph{not} coincide with density-derivatives of the thermodynamic entropy density $s[\mathbf{q}]=\sum_{a}\mu_{a}q_{a}-f[\boldsymbol{\mu}]$ in the microscopic model.

To enforce stationarity of $\mathcal{P}_{\rm eq}$ under $\mathcal{J}^{\rm R}_{a}[\boldsymbol{\psi}]$, we must mandate the rate of entropy production under the conservative current vanishes:
\begin{equation}
    \label{eq:Liouville_orthogonality}
    \int dx\,\sigma^{\rm R}_{S}(x)\equiv 
    \sum_{a}\int dx\,\mathcal{J}^{\rm R}_{a}[\boldsymbol{\gamma}(x)]\partial_{x}\gamma_{a}(x) = 0.
\end{equation}
Here $\gamma_{a}(x)$ represent nonlinear thermodynamic forces conjugate to $\psi_{a}(x)$,
\begin{equation}
    \label{eq:gamma}
    \gamma_{a}[\boldsymbol{\psi}]\equiv \frac{\delta \mathcal{S}_{\rm eq}[\boldsymbol{\psi}(x)]}{\delta \psi_{a}(x)}.
\end{equation}
As detailed out in End Matter, from the time evolution of $\mathscr{S}(x)$ one can infer that $\sigma^{\rm R}_{S}(x)=\partial_{x}\mathscr{J}^{\rm R}_{S}(x)$, where $\mathscr{J}^{\rm R}_{S}(x)=\sum_{a}\mathcal{J}^{\rm R}_{a}[\boldsymbol{\gamma}]\gamma_{a}(x)$ is the entropy current density under reversible flow.

To manifestly satisfy stationary under the reversible flow, Eq.~\eqref{eq:Liouville_orthogonality}, we introduce the \emph{master potential} $\mathcal{G}[\boldsymbol{\gamma}]$ via
\begin{equation}
    \mathcal{J}^{\rm R}_{a}[\boldsymbol{\gamma}(x)] = \frac{\delta \mathcal{G}[\boldsymbol{\gamma}]}{\delta \gamma_{a}(x)},
\end{equation}
where $\mathcal{G}[\boldsymbol{\gamma}]=\int dx\,\mathscr{G}[\boldsymbol{\gamma}]$ is in general a non-linear functional of $\boldsymbol{\gamma}$-fields,
\begin{equation}
    \label{eq:master_potential}
    \mathscr{G}[\boldsymbol{\gamma}] = \sum_{n\geq 0}\frac{1}{n!}
    \sum_{j_{1},\ldots,j_{n}}G^{\flat}_{j_{1},\ldots,j_{n}}\prod_{i=1}^{n}\gamma_{j_{i}}(x),
\end{equation}
specified by completely symmetric tensors $G^{\flat}_{j_{1},\ldots,j_{n}}$ (more generally, $\mathcal{G}[\boldsymbol{\gamma}]$ can also include gradients of $\boldsymbol{\gamma}$-fields, in which case $G^{\flat}_{j_{1},\ldots,j_{n}}$ are promoted to differential operators). %For a master potential of the form \eqref{eq:master_potential}, one has $\mathcal{V}[\boldsymbol{\gamma}]=\mathscr{G}[\boldsymbol{\gamma}]$.
Indeed, $\mathscr{G}[\boldsymbol{\gamma}]$ is none other than the EFT counterpart of the free-energy flux $g(\boldsymbol{\mu})$: the Maxwell relation for \emph{dressed} tensors, namely $B^{\rm dr}_{a|b}=B^{\rm dr}_{b|a}$ for $B^{\rm dr}_{a|b}\equiv \int dx \langle \mathcal{J}_{a}(x)\psi_{b}(0) \rangle^{c}_{\mathcal{P}_{\rm eq},\boldsymbol{\xi}}$, is a direct corollary of stationarity and translational invariance, cf. End Matter.
%An alternative derivation based on the Fokker-Planck formalism is presented in \cite{SM}.

Finally, we introduce the dissipative component of the hydrodynamic current in the form
\begin{equation}
    \mathcal{J}^{\rm D}_{a}[\boldsymbol{\psi}] = -\sum_{b}D^{\flat}_{ab}\partial_{x}\gamma_{b}(x),
\end{equation}
where $D^{\flat}\geq 0$ is required for the positive production of entropy density, $\sigma_{S}(x)=\sum_{a,b}(\partial_{x}\gamma_{a})D^{\flat}_{ab}(\partial_{x}\gamma_{b})$.
Note that $\mathcal{J}^{\rm D}_{a}[\boldsymbol{\psi}]$ cannot preserve $\mathcal{P}_{\rm eq}$ on its own, but instead has to be counterbalanced by suitable stochastic (Gaussian) fields $\xi_{a}(x)$ with $\langle \xi_{a}(x,t)\rangle_{\boldsymbol{\xi}}=0$ and covariance
$\langle \xi_{a}(x,t)\xi_{b}(x',t')\rangle^{c}_{\boldsymbol{\xi}} = \sigma^{\flat}_{ab}\delta(x-x')\delta(t-t')$ set by FDR, $\sigma^{\flat}_{ab}=2D^{\flat}_{ab}$. More generally, when $D^{\flat}[\boldsymbol{\psi}]$ can depend on $\boldsymbol{\psi}$-fields, leading to multiplicative noise.

%In summary, by specifying $A^{\flat}_{(n)}$ an input, we first determine $K_{(n)}$ from Eq.~\eqref{eq:Liouville_orthogonality}, and subsequently, by employing the associated conjugate fields $\gamma_{a}(x)$, construct $\mathcal{J}^{\rm D}_{a}[\boldsymbol{\psi}]$ by prescribing the bare Onsager matrix $L^{\flat}$ (while adjusting the noise covariance $\sigma^{\flat}$).

\paragraph*{\textbf{Minimal models of nonlinear EFTs.}}
To benchmark our framework, we consider a family of \emph{exactly truncated} `minimal models' comprising of two interacting chiral modes $\boldsymbol{\psi}=\{\psi_1,\psi_2\}$, one exhibiting diffusive and the other superdiffusive broadening.
We consider a \emph{quartic}, non-Gaussian, measure $\mathcal{S}_{\rm eq}[\boldsymbol{\psi}]=\int\mathrm{d}x\,\mathscr{S}[\boldsymbol{\psi}]$, of the form
\begin{equation}
    \label{eq:model_measure}
    \mathscr{S} = \frac{1}{2}(\psi_1^2+\psi_2^2)
    -\frac{g}{6c}\psi_1^3 +
    \frac{\chi g}{2c}\psi_1\psi_2^2
    +\frac{\kappa}{24}(\psi_1^4+\psi_2^4).
\end{equation}
\noindent
with free coupling constants $g,\chi,c,\kappa>0$. Picking a linear master potential $\mathscr{G}[\boldsymbol{\gamma}]=-\tfrac{1}{2}c(\gamma^{2}_{1}-\gamma^{2}_{2})$, we generate the following reversible current components
\begin{equation}
    \label{eq:model_currents}
    \begin{split}
    &\mathcal{J}^{\rm R}_1 =
    -c\,\psi_1+\frac{g}{2}\psi_1^2-\frac{\chi\,g}{2}\psi_2^2
    -\frac{c\,\kappa}{6}\psi_1^3,
    \\
    &\mathcal{J}^{\rm R}_2=
    c\,\psi_2+\chi\,g \psi_1\psi_2 +\frac{c\,\kappa}{6}\psi_2^3.
    \end{split}
\end{equation}
We note that stationarity conditions are satisfied at both bare and dressed levels by design. Finally, we pick a diagonal form for the dissipative current component by setting $L_{ab}^{\flat}=\delta_{ab} D_a^{\flat}$, implying
\begin{equation}
    \label{eq:model_dissipation}
    \mathcal{J}_a^{\rm D} = - D_a^{\flat} \partial_x \gamma_a,
\end{equation}
\noindent
and adjust the covariance of noise fields $\xi_a(x,t)$ setting $\sigma^{\flat}_{ab}=2\delta_{ab}D_a^\flat$. Unlike in theories with Gaussian stationary states, the dissipative component $\mathcal{J}^{\rm D}_{a}[\boldsymbol{\psi}]$ features \emph{nonlinear} terms in $\boldsymbol{\psi}$-fields.

\begin{figure}[t!]
    \centering
\includegraphics[width=0.5\textwidth]{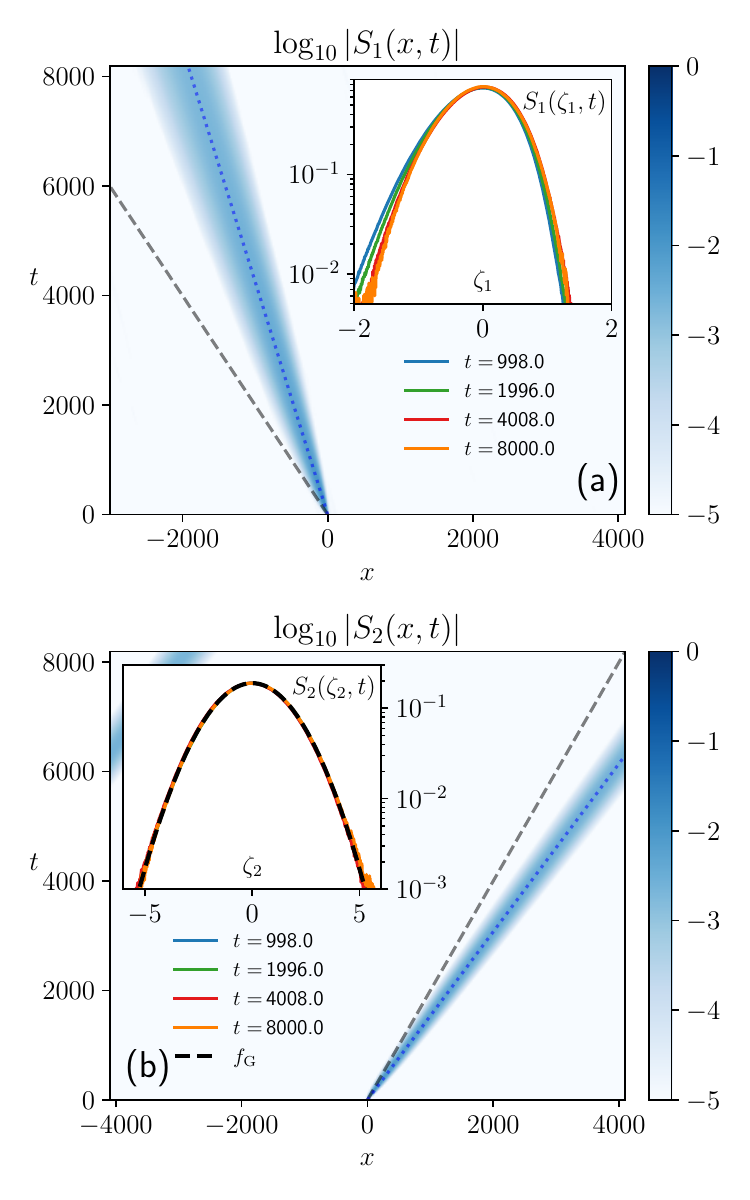}
    \caption{\textbf{Diagonal structure factors in a minimal model of a two-component nonlinear EFT.} Time evolution of $S_{a}(x,t)$ in the minimal model (cf. Eqs.~\eqref{eq:model_measure}, \eqref{eq:model_currents} and \eqref{eq:model_dissipation}) with bare parameters $c=0.5$, $g=1.0$, $\chi=0.1$ and $\kappa=1.0$. Gray dashed lines indicate the bare characteristics $v^{\flat} t$ with $v^{\flat}=\pm c$, while blue dotted lines trace the actual peak trajectories $v_{a}t$ propagating with dressed velocities ($v_1\approx -0.234, v_2\approx0.650$). Insets demonstrate collapse of the rescaled structure factors $  t^{1/z_a}S_{a}\left(\zeta_{a},t\right)$, with scaling variables
    $\zeta_{a}(x,t)\equiv t^{-1/z_a}(x-v_{a}t)$. \textbf{(a)} Mode 1 exhibits superdiffusive broadening with $z_1=3/2$. \textbf{(b)} Mode 2 exhibits diffusive transport with $z_2=2$ and a Gaussian scaling function (black dashed line). Simulation parameters: $N=L=8192$, $\Delta t=0.04$, number of samples $2\times10^4$.}
    \label{fig:struct_factors}
\end{figure}

We consider the diagonal dynamical structure factors $S_{a}(x,t)\equiv \langle \phi_{a}(x,t)\phi_{a}(0,0) \rangle^{c}_{\mathcal{P}_{\rm eq},\boldsymbol{\xi}}$ of EFT normal modes $\phi_{a}(x,t)\equiv \sum_{b}R_{ab}\psi_{b}(x,t)$, where $R$ denote the eigenvectors of the dressed current Jacobian $A^{\rm dr}\equiv B^{\rm dr}[C^{\rm dr}]^{-1}$, such that $RC^{\rm dr}R^{\rm T}=1$, with $C^{\rm dr}_{ab}=\int dx\langle \psi_{a}(x)\psi_{b}(0) \rangle^{ c}_{\mathcal{P}_{\rm eq}}$. Dressed velocities $v_{a}$ appear as eigenvalues of $A^{\rm dr}$.

The results of our numerical simulation, showing the dynamical structure factors $S_{a}(x,t)$
and the corresponding scaling behavior $S_{a}(x,t) \asymp (\lambda_{a}t)^{-1/z_a}\mathscr{F}_a\left((x-v_at)(\lambda_{a}t)^{-1/z_a} \right)$ with dynamical exponents $z_{a}$ and scaling functions $\mathscr{F}_{a}$, are summarized in Fig.\,\ref{fig:struct_factors} (see End Matter for details): the first mode ($a=1$) is superdiffusive, with exponent $z_{1}=3/2$, as expected due to its self-coupling.
The second normal mode ($a=2$) is diffusive, with $z_2=2$ and a Gaussian scaling function $\mathscr{F}_{\rm G}(\zeta_{2})=(2\pi)^{-1/2}e^{-\zeta^{2}_{2}/2}$ with $\zeta_{2}=(x-v_2t)/\sqrt{4D_{2}t}$; the propagation velocities $v_{a}$ of normal modes $\phi_{a}(x,t)$ differ from bare velocities $v^{\flat}_{a}\in \{-c,+c\}$, whereas the fitted value $D_{2} \approx 1.32$ differs from the bare diffusion constant $D^\flat_{2} = 1$ (see End Matter for additional details).

\paragraph*{\textbf{Conclusion.}}
In this work, we presented a general theoretical framework for constructing effective field theories of hydrodynamic systems in one spatial dimension. In our approach, the master potential takes a central role since it manifestly implements the Maxwell relation at the EFT level, a key ingredient missed in the conventional approach.

We implemented a numerical integration scheme for solving coupled stochastic differential equations with nonlinear currents and non-Gaussian stationary states, and benchmarked it on a simple system of two interacting chiral modes that features anomalous, superdiffusive transport.

To the best of our knowledge, our work constitutes the first successful numerical solution of nonlinear fluctuating hydrodynamics with non-Gaussian equilibrium states. This is achieved with the aid of a particular discretization procedure that ensures stationarity of non-Gaussian measures (detailed out in Supplemental Material \cite{SM}),
in contrast to previously employed lattice discretizations which only apply to Gaussian states \cite{PhysRevLett.131.197102,roy2024universality, Schmidt_2025, Schmidt_2026, minami2026symmetry}, which thus have a limited range of applications. The presented approach permits a fully general structure of coupling tensors and therefore enables studying transport phenomena in generic one-dimensional (deterministic or stochastic) interacting systems.

The natural next step is to implement our computational scheme in specific, physically relevant, conservative systems such as, for example, the FPUT chains. To accomplish this, it is imperative to devise systematic truncation schemes. The final computational step is then to relate the dressed EFT tensors to equilibrium static correlators of the target microscopic model by numerically solving the `inversion problem' \cite{in_prep}, modulo systematic errors due to truncation and discretization.
Such an approach would already suffice to reproduce the correct leading asymptotic behavior of dynamical structure factors.

Our computational framework permits us to systematically test the predictions of the standard mode-coupling theory \cite{popkov2015fibonacci,popkov2016exact}, and to study more interesting cases of umbilic systems \cite{koterle2026anomalous,Schmidt_2026,Paljk_2026} with non-Gaussian states. On the other hand, there are several possible extensions, including e.g. generalization to higher spatial dimensions, subdiffusive systems \cite{Feldmeier_2020,Gromov_2020,Meerson_2024,Prelovsek_2024,SciPostPhys.20.5.145}, and generalized hydrodynamics \cite{Castro_Alvaredo_2016,Bertini_GHD,Bastianello_2018,Doyon_2020,Bastianello_2024}, to mention a few.

\paragraph*{Acknowledgements.}
We thank F. H\" ubner, \v{Z}. Krajnik, P. Orlov, L. Paljk, R. Sharipov and T. Yoshimura for useful remarks and discussions. EI is supported by the Research Program P1-0402 and Project N1-0368 funded by the Slovenian Research Agency (ARIS). MK acknowledges support by ERC Advanced grant No.~101096208 -- QUEST, and Research Program P1-0402 and Grant N1-0368 of Slovenian Research and Innovation Agency (ARIS).

\bibliography{refs}

\section*{End Matter}

\subsection{Mode-coupling equations}

The present theoretical understanding of transport phenomena in strongly interacting one-dimensional systems supporting ballistic (i.e. sound) hydrodynamic modes essentially rests on the mode-coupling theory~\cite{van_Beijeren_2012}. The standard mode-coupling equations (MCE) constitute a system of \emph{deterministic} partial differential equations for diagonal dynamical structure factors $S_{a}(x,t)\equiv \langle \phi_{a}(x,t)\phi_{a}(0,0) \rangle^{c}_{\mathcal{P}_{\rm eq},\boldsymbol{\xi}}$ of the form~\cite{spohn2014nonlinear}
\begin{equation}
    \label{eq:MCE}
    \big(\partial_{t}+\mathrm{v}_{a}\partial_{x}-\mathrm{D}_{a}\partial^{2}_{x}\big)S_{a}(x,t) = [S_{a}\star M_{a}](x,t),
\end{equation}
where $S_{a}\star M_{a}\equiv \int^{t}_{0}\mathrm{d}\tau \int \mathrm{d}y \, S_{a}(x-y,t-\tau)\partial_{y}^{2}M_{a}(y,\tau)$, and $M_{a}(y,\tau)=\sum_{b,c}[\mathrm{G}^{a}_{bc}]^{2}S_{b}(y,\tau)S_{c}(y,\tau)$ is the so-called memory kernel. By applying the Mori--Zwanzig projection on nonlinear fluctuating hydrodynamics, Ref.~\cite{spohn2014nonlinear} derives Eqs.~\eqref{eq:MCE} by assuming (a) Gaussian stationary measure with \emph{completely symmetric} $\rm{G}$, (b) nonlinear hydrodynamic currents $\mathcal{J}^{\rm R}_{a}[\boldsymbol{\phi}]$ of the form \eqref{eq:NLFHD_currents}, and (c) replacing the bare propagator with the fully dressed one.
Hence, equations \eqref{eq:MCE} are merely an \emph{approximation} to $S_{a}(x,t)$. Indeed, the asymptotic scaling functions $\mathscr{F}_{a}$ of $S_{a}(x,t)$ with non-vanishing self-couplings, $\mathrm{G}^{a}_{aa}\neq 0$, do not match the universal Pr\"{a}hofer--Spohn distribution of the KPZ equation. We also note that phenomenological inclusion of \emph{bare} diffusion coefficients ${\rm D}_{a}$ precludes quantitative predictions of \emph{dressed} diffusion constants of diffusive, Gaussian peaks
(see Ref.~\cite{convective} for a detailed discussion).

In strictly hyperbolic systems (i.e. for mutually distinct sound-mode velocities), the asymptotic solutions to MCE \eqref{eq:MCE} yield an infinite tower of `universality classes' depending on the structure of $\mathrm{G}$, characterized by superdiffusive algebraic exponents $z_{a} \in [3/2,2]$ and universal scaling function \cite{popkov2015fibonacci,popkov2016exact}. Remarkably, in the absence of self-couplings, the mode-coupling equations \eqref{eq:MCE} provide quantitatively exact predictions for the scaling functions \cite{popkov2015fibonacci}. Crucially, however, Refs.~\cite{popkov2015fibonacci,popkov2016exact} utilize the mode-coupling equations for the dynamical correlation functions of physical charge densities in the full, \emph{microscopic} theory (with \emph{unrestricted} tensor ${\rm G}$ computed from equilibrium averages of microscopic current densities), as opposed to Eqs.~\eqref{eq:MCE} obtained in the scope of the stochastic field theory of nonlinear fluctuating hydrodynamics.
In the conventional setting described by Eqs.~\eqref{eq:NLFHD}, with quadratic currents \eqref{eq:NLFHD_currents} and Gaussian stationary measures $\mathcal{P}_{\rm eq}$, the coupling tensor $\rm{G}$ must be completely symmetric, and thus the corresponding mode-coupling equations lose predictive power. On the other hand, upon relaxing this constraint on the EFT side, one is inevitably confronted with higher-order nonlinearities and non-Gaussian measures that preclude the application of Wick theorem.
Justifying the mode-coupling equations in a controlled setting of consistent hydrodynamic EFT therefore remains an open question.

\subsection*{Hydrodynamic entropy}
In terms of nonlinear thermodynamic forces $\gamma_{a}(x)$ (cf. Eq.~\eqref{eq:gamma}), the hydrodynamic entropy density $\mathscr{S}[\boldsymbol{\psi}]$ satisfies
\begin{equation}
    \partial_{t}\mathscr{S}(x,t) + \sum_{a}\gamma_{a}(x)\partial_{x}\mathcal{J}_{a}[\boldsymbol{\gamma}] = 0.
\end{equation}
This can be recast in the form of a local continuity equation with a source,
\begin{equation}
    \partial_{t}\mathscr{S}(x,t)+\partial_{x}\mathscr{J}_{\rm S}(x,t)=\sigma_{S}(x,t),    
\end{equation}
where $\mathcal{J}_{\rm S}[\boldsymbol{\gamma}]=\sum_{a}\mathcal{J}_{a}[\boldsymbol{\gamma}]\gamma_{a}$ and $\sigma_{\rm S}[\boldsymbol{\gamma}]=\sum_{a}\mathcal{J}_{a}[\boldsymbol{\gamma}]\partial_{x}\gamma_{a}$,
are the total entropy current density and entropy production rate, respectively.

\subsection*{Maxwell relation in hydrodynamic EFTs}
Using stationarity and translational symmetry, we here establish that $B^{\rm dr}_{a|b}=\lim_{L\to \infty} L^{-1}\int\mathrm{d}x\mathrm{d}y\, \mathcal{B}_{a|b}(x,y)$, with
\begin{equation}
    \mathcal{B}_{a|b}(x,y) \equiv \langle \mathcal{J}_{a}(x)\psi_{b}(y) \rangle^{c}_{\mathcal{P}_{\rm eq},\boldsymbol{\xi}},
\end{equation}
is a \emph{symmetric} tensor, namely $B^{\rm dr}_{a|b}=B^{\rm dr}_{b|a}$.
We first show that it suffices to only consider the reversible component $\mathcal{J}^{\rm R}_{a}[\boldsymbol{\psi}]$ in $\mathcal{J}_{a}[\boldsymbol{\psi}]$: stochastic current is uncorrelated with $\boldsymbol{\psi}$-fields and thus its contribution vanishes upon averaging over noise; the dissipative component $\mathcal{J}^{\rm D}_{a}[\boldsymbol{\psi}]$ on the other hand can be cast as a total derivative,
\begin{equation}
    -\sum_c D_{ac}^\flat \int \mathrm{d}x \, \partial_x \langle \gamma_c(x) \psi_b(y) \rangle^{c}_{\mathcal{P}_{\mathrm{eq}}}=0.
\end{equation}
Subsequently, we thus focus on $\mathcal{B}^{\rm R}_{ab}(x,y)\equiv \langle \mathcal{J}^{\mathrm{R}}_a(x) \psi_b(y)\rangle^{c}_{\mathcal{P}_{\mathrm{eq}}}$.
Since the equilibrium measure is stationary under the temporal evolution generated by the reversible current $\mathcal{J}^{\rm R}_{a}[\boldsymbol{\psi}]$, $\partial_t^R \psi_a = -\partial_x \mathcal{J}_a^R$, we have
\begin{equation}
    \begin{split}
    &\partial_t^R \langle \psi_a(x) \psi_b(y)\rangle_{\mathcal{P}_\mathrm{eq}}^c
    = \left[- \partial_x \mathcal{B}^R_{ab}(x,y) 
       - \partial_y \mathcal{B}^R_{ba}(y,x) \right]\\
    &\quad = -\partial_r \left[ \mathcal{B}_{ab}^R(r,0) - \mathcal{B}_{ba}^R(-r, 0) \right]=0,
    \end{split}
\end{equation}
\noindent
where we used the relative coordinate $r\equiv x-y$. Therefore $\mathcal{B}_{ab}^R(r) = \mathcal{B}_{ba}^R(-r)$, and, assuming clustering, it follows
\begin{equation}
    B_{ab}^{\mathrm{dr}} = \int \mathrm{d}r \mathcal{B}_{ab}^R(r) =
    \int \mathrm{d}r \, \mathcal{B}_{ba}^R(-r) = B_{ba}^{\mathrm{dr}}.
\end{equation}

\subsection*{Numerical implementation}
We consider a system of length $L$, impose periodic boundary conditions, and discretize space into $N$ equal segments of length $\Delta x$: we set $L=N$ or, equivalently, lattice spacing $\Delta x=1$. The fields in discrete space are represented as $\psi_a(x,t)=\psi_a(j\Delta x, t)\equiv\psi_{a,j}(t)$, with lattice sites $j\in\{1,\ldots,N\}$. Unless explicitly needed, we suppress writing the time dependence.

In the lattice model, the ultralocal stationary measure $\mathcal{S}_{\rm eq} = \int \mathrm{d}x\,\mathscr{S}(x)$, factorizes into one-site measures, $\prod_{j=1}^N \exp[-\Delta x \mathscr{S}_j]$,
where $\mathscr{S}_j\equiv \mathscr{S}(j\Delta x)$, which can be efficiently sampled using the Metropolis algorithm. To enforce the constraints $\langle \psi_{a,j}\rangle=0$ for finite $L$,
we tilt the measure by including chemical potentials $h_a$,
$\mathscr{S}_j \to \mathscr{S}_j + \sum_{a} h_a \psi_{a,j}$,
and determine $h_a$ variationally using the Newton iteration.
Our implementation of the Metropolis algorithm is detailed in \cite{SM}.

Introducing discrete current densities $\mathcal{J}_{a,j+1/2}$ positioned at the interfaces between lattice sites, the stochastic Langevin equation of the form \eqref{eq:NLFHD} can be discretized in a flux-preserving form $\partial_t \psi_{a,j}=- (\mathcal{J}_{a,j+1/2}-\mathcal{J}_{a,j-1/2})/\Delta x$.
The currents are further split into separate reversible, dissipative and stochastic parts. Time evolution is carried out using a third-order strong-stability-preserving stochastic Runge-Kutta scheme with a fixed timestep $\Delta t$ (the same general approach is used in Refs.\, \cite{garcia2024introduction,srivastava2023staggered, minami2026symmetry}), see \cite{SM} for additional details on the implementation. 

After spatial discretization, the stochastic difference-differential equations do not generically preserve the discretized stationary measure of the continuum model \eqref{eq:NLFHD}. Finding an exact, measure-preserving, discretization is a challenging problem that delicately depends on the form of continuum currents. While the measure inevitably ceases to be preserved upon discretizing time, this can be simply controlled by reducing the timestep $\Delta t$.

The strategy we used was the following: by prescribing a discretization scheme, we probed deviations from stationarity by monitoring the equal-time $n-$point correlation functions (written here for the field theory),
\begin{equation}
    \begin{split}&
    \mathscr{C}_{(n)}\equiv \mathscr{C}_{i_1\ldots i_n}(0,r_2\ldots r_n,t) =\\
    &=\langle \psi_{i_1}(x,t)\psi_{i_2}(x+r_2,t)\ldots\psi_{i_n}(x+r_n)\rangle^c_{\mathcal{P}_{\rm eq}, \boldsymbol{\xi}}
    \end{split},
\end{equation}
\noindent
Since the equilibrium measure \eqref{eq:equilibrium_entropy} is local, we require the tensors $\mathscr{C}_{(n)}$ to asymptotically become  of `contact form',
\begin{equation}
\lim_{t\to\infty}\mathscr{C}_{a_1\ldots a_n}(0,r_2\ldots r_n,t)= M_{a_1\ldots a_n}\prod_{m=2}^n \delta(r_m),
\end{equation}
where
$M_{(n)}$ are equilibrium equal-space correlators given by
\begin{equation}
    M_{(n)} \equiv M_{a_1\ldots a_n} = \frac{1}{L}\int\mathrm{d}x\langle \psi_{a_1}(x)\ldots\psi_{a_n}(x)\rangle^c_{\mathcal{P}_{\rm eq}}.
\end{equation}
Validating that the (lattice) measure $\mathcal{P}_{\rm eq}$ is stationary requires: (i) showing that $\mathscr{C}_{(n)}$ attain the contact form, and (ii) showing $M_{(n)} = \mathcal{M}_{(n)}(t)$ for all times $t$, where
\begin{equation}
     \mathcal{M}_{(n)} \equiv \mathcal{M}_{a_1 \ldots a_n}(t) = \frac{1}{L}\int\mathrm{d}x\,\big\langle \prod_{i=1}^{n}\psi_{a_i}(x,t)\big\rangle^{c}_{\mathcal{P}_{\rm eq},\boldsymbol{\xi}}.
\end{equation}

Numerical computation of structure factors $S_{a}(x,t)$ requires averaging over many initial samples drawn from the discrete measure $\mathcal{P}_{\rm eq}$. By exploiting stationarity, we additionally average over all origins and time shifts (see SM \cite{SM}).

\begin{figure}[h!]
    \centering
\includegraphics[width=0.5\textwidth]{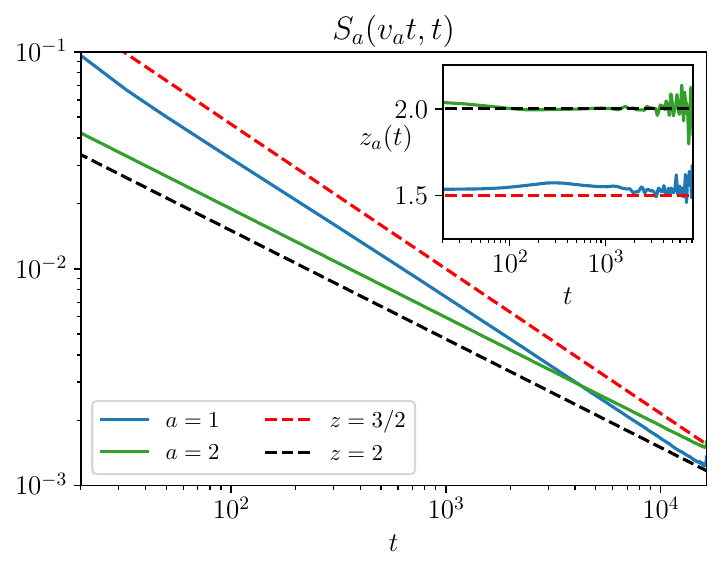}
    \caption{\textbf{Decay of diagonal structure factors and running exponents.} The decay of the peaks of diagonal structure factors $S_a(v_at,t)$ in the minimal model. The dashed lines show the expected asymptotic decay rates $t^{-1/z_{a}}$ for $z_{a}\in\{3/2, 2\}$. In the inset we show time-resolved dynamical exponents $z_a(t)=-[{\rm d}\!\log S_{a}(v_a t,t)/\mathrm{d}\log \! t]^{-1}$ which shows that mode 1 is superdiffusive with $z_1 (t)\approx 3/2$, while mode 2 clearly converges to the expected value $z_2 \approx 2$. 
    Parameters of simulation: $N=L=8192$, $\Delta t=0.04$, number of samples $2\times10^4$.}
    \label{fig:running_exponents}
\end{figure}

Since algebraic exponents $z_{a}$ and scaling functions $\mathscr{F}_{a}$ are defined as asymptotic objects, it is more customary to compute the running exponents $z_a(t)=-[{\rm d}\!\log S_{a}(v_a t,t)/\mathrm{d}\log \! t]^{-1}$ and track their convergence with time, as exemplified in Fig.\,\ref{fig:running_exponents} for the minimal model introduced in the text. We note that $z_{a}(t)$ are additionally smoothed using  a weighted Gaussian averaging over each $5$ consecutive points with variance $\sigma=5$. As shown in Fig.~\ref{fig:running_exponents}, on the timescale of the simulation, both running exponents approach their expected asmyptotic values $z_1 (t)\approx3/2$ and $z_2 (t) \approx 2$, respectively, compatibly with the rescaling used in the main text.

\clearpage

\onecolumngrid
\appendix
\setcounter{secnumdepth}{2}

\begin{center}
	{\Large Supplemental Material - Effective field theories of nonlinear fluctuating hydrodynamics in one dimension}
\end{center}
\begin{center}
	{Matija Koterle, Enej Ilievski}
\end{center}

The supplemental material discusses the following technicalities in detail:
\begin{itemize}
    \item Appendix \ref{appendix:stationarity} shows the derivation of the stationarity condition (Maxwell relation) starting from the Fokker-Planck equation and discusses minimal models which, as expansions in terms of the thermodynamic force, satisfy this condition automatically.
    \item Appendix \ref{appendix:equilibration} discusses the equilibration to the ultra-local equilibrium ensemble under stochastic dynamics.
    \item Appendix \ref{appendix:numerics} details the numerical implementation of the integration scheme. It compares the scheme used for models with a cyclic Hessian and Gaussian equilibrium measures with consistent schemes which include nonlinear damping.
    \item Appendix \ref{appendix:metropolis} includes the implementation of the Metropolis algorithm for sampling equilibrium states $\mathcal{S}_{\rm eq}$ as well as the Newton iteration.
\end{itemize}

%\tableofcontents

\section{Stationarity conditions}\label{appendix:stationarity}

\subsection{Fokker-Planck formalism}
Given the equation of motion for the fields $\boldsymbol{\psi}=\{\psi_1(x),\ldots\psi_{N_Q}(x)\}$
and the ultra-local equilibrium measure $\mathcal{P}_{\mathrm{eq}}$
\begin{equation}
    \partial_t \psi_a
    +
    \partial_x
        \mathcal{J}_a[\boldsymbol{\psi}]
    =
    0,
    \quad
    \mathcal{P}_{\mathrm{eq}}[\boldsymbol{\psi}]
    =
    Z^{-1} \exp\left(-\mathcal{S}_{\rm eq}[\boldsymbol{\psi}]\right),
    \quad 
    \mathcal{S}_{\rm eq}[\boldsymbol{\psi}]
    =
    \int \mathrm{d}x\, \mathscr{S}[\boldsymbol{\psi}],
\end{equation}
\noindent
the currents are decomposed into reversible (non entropy producing), dissipative (entropy producing) and noise parts $\mathcal{J}_a^{\rm R}[\boldsymbol{\psi}]
+\mathcal{J}_a^{\rm D}[\boldsymbol{\psi}] + \xi_a$.
We represent the reversible part as an expansion in nonlinearities, while the dissipative
term is written as the gradient flow of the same action $\mathcal{S}_{\rm eq}$,
\begin{equation}
    \mathcal{J}_a^{\rm R} [\boldsymbol{\psi}] = 
    \sum_{n\geq 1}
    \frac{1}{n!}
    \sum_{b_1 \ldots b_n}
    A_{a| b_1 \ldots b_n}^\flat
    \psi_{b_1}
    \ldots \psi_{b_n},
    \quad
    \mathcal{J}_a^{\rm D} [\boldsymbol{\psi}] =
    - \sum_{b} D_{ab}^\flat \partial_x \frac{\delta \mathcal{S}_{\rm eq}}{\delta \psi_b}
    ,
\end{equation}
\noindent
with white Gaussian noise $\langle \xi_a(x,t)\xi_b(y,s)\rangle_{\mathbf{\xi}}
=\sigma_{ab}\delta(x-y)\delta(t-s)$ with zero-mean $\langle\xi_a(x,t)\rangle_{\mathbf{\xi}}=0$.
Since the action is ultra-local, $\delta \mathcal{S}_{\rm eq}/\delta \psi_a =
\partial \mathscr{S}/\partial \psi_a$. The associated Fokker-Planck equation for 
the time-dependent distribution $\mathcal{P}_t[\boldsymbol{\psi}]$ is \cite{minami2026symmetry}
\begin{equation}
    \partial_t \mathcal{P}_t = -\sum_a \int \mathrm{d}x 
    \left(\partial_x \frac{\delta}{\delta \psi_a}\right)
    \left[
        \mathcal{J}_a^{\rm R} - \sum_b 
        D_{ab}^\flat \partial_x \frac{\delta \mathcal{S}_{\rm eq}}{\delta \psi_b}
        - \sum_b \frac{1}{2}\sigma_{ab} \partial_x \frac{\delta}{\delta \psi_b} 
    \right] \mathcal{P}_t .
    \label{eq:supp-functional-fokker-planck}
\end{equation}
The stationary (equilibrium)  measure is defined as $\partial_t \mathcal{P}_{\mathrm eq} = 0$.
For the proposed distribution above, we have
\begin{equation}
    \frac{\delta \mathcal{P}_{\mathrm{eq}}}{\delta \psi_a}
    =
    - \frac{\delta \mathcal{S}_{\rm eq}}{\delta \psi_a} \mathcal{P}_{\mathrm{eq}},
\end{equation}
\noindent
which makes the dissipative and noisy parts cancel point-wise
\begin{equation}
    \left[-\sum_b D_{ab}^\flat \partial_x \frac{\delta \mathcal{S}_{\rm eq}}{\delta \psi_b} - 
    \sum_{b} \frac{1}{2} \sigma_{ab} \partial_x \frac{\delta}{\delta \psi_b}\right]\mathcal{P}_{\mathrm{eq}} =
    \left[-\sum_{b} D_{ab}^\flat \partial_x \frac{\delta \mathcal{S}_{\rm eq}}{\delta \psi_b} +
    \sum_b \frac{1}{2} \sigma_{ab} \partial_x \frac{\delta \mathcal{S}_{\rm eq}}{\delta \psi_b}\right]\mathcal{P}_{\mathrm{eq}} = 0,
\end{equation}
\noindent
if $\sigma_{ab}=2D^\flat_{ab}$. This is the detailed-balance, or fluctuation-dissipation, part of the
construction.

We are left with the stationarity of the reversible part of the evolution. 
It is convenient to express it in the following way by use of an integration
by parts
\begin{equation}
    \sum_a\int \mathrm{d}x\,
    \left(\partial_x\frac{\delta}{\delta\psi_a}\right)
    \left[\mathcal{J}_a^{\rm R} (x)\mathcal{P}_{\mathrm{eq}}\right]
    =
    -\sum_a\int \mathrm{d}x\,
    \frac{\delta}{\delta\psi_a}
    \left[
        \partial_x \mathcal{J}_a^{\rm R}(x)\mathcal{P}_{\mathrm{eq}}
    \right],
\end{equation}
where $\partial_x\mathcal{P}_{\mathrm{eq}}[\boldsymbol{\psi}]=0$ because
$\mathcal{P}_{\mathrm{eq}}$ has no free spatial label. We separately analyse each of the
two terms. The first is the derivative of the current,
giving the phase-space divergence,
\begin{equation}
    \left[
    \sum_a\int \mathrm{d}x\,
    \frac{\delta\,\partial_x \mathcal{J}_a^{\rm R}}
    {\delta\psi_a}
    \right]\mathcal{P}_{\mathrm{eq}}
    \sim
    \left[
    \sum_{a}\int \mathrm{d}x\,
    \partial_x
    \frac{\partial \mathcal{J}_a^{\rm R}}{\partial\psi_a}
    \right]\mathcal{P}_{\mathrm{eq}}
    =
    0.
\end{equation}
While the functional derivative acts on $\mathcal{J}_a^{\rm R}$, this contribution
vanishes because it is a spatial total derivative on a periodic domain.
The remaining term is
\begin{equation}
    -\sum_a\int \mathrm{d}x\,
    \partial_x \mathcal{J}_a^{\rm R}
    \frac{\delta \mathcal{S}_{\rm eq}}{\delta\psi_a}
    \mathcal{P}_{\mathrm{eq}}
    =
    \sum_a\int \mathrm{d}x\,
    \mathcal{J}_a^{\rm R}
    \partial_x\frac{\delta \mathcal{S}_{\rm eq}}{\delta\psi_a}
    \mathcal{P}_{\mathrm{eq}},
\end{equation}
\noindent
where the partial derivative was moved by integrating by parts. Stationarity 
of the reversible part therefore requires the nontrivial condition
\begin{equation} \label{eq:deterministic_reversibility}
    \sum_a \int \mathrm{d}x\, \mathcal{J}^{\rm R}_a \partial_x \frac{\delta \mathcal{S}_{\rm eq}}{\delta \psi_a}=0,
\end{equation}
\noindent
for arbitrary periodic fields. Since $\mathcal{S}_{\rm eq} = \int \mathrm{d}y\, \mathscr{S}(y)$ then
\begin{equation}
    \frac{\delta \mathcal{S}_{\rm eq}}{\delta \psi_a} = 
    \sum_i \int \mathrm{d}y \frac{\partial \mathscr{S}(\boldsymbol{\psi}(y))}{\partial \psi_i(y)} 
    \frac{\delta \psi_i(y)}{\delta \psi_a(x)}=
    \sum_a \int \mathrm{d}y \frac{\partial \mathscr{S}(\boldsymbol{\psi}(y))}{\partial \psi_i(y)} 
    \delta_{ia}\delta(x-y)
    =
    \frac{\partial \mathscr{S}(\boldsymbol{\psi}(x))}{\partial \psi_a(x)} 
\end{equation}
Then we can use the chain rule to obtain
\begin{equation}
    \partial_x \frac{\delta \mathcal{S}_{\rm eq}}{\delta \psi_a} = 
    \partial_x \frac{\partial \mathscr{S}}{\partial \psi_a} = 
    \sum_b \left(\partial_{\psi_a} \partial_{\psi_b} \mathscr{S} \right) \partial_x \psi_b,
\end{equation}
Next we insert this into Eq.\,\eqref{eq:deterministic_reversibility} 
\begin{equation}
    \int \mathrm{d}x\,\sum_a \mathcal{J}_a^{\rm R}
    \partial_x \frac{\delta \mathcal{S}_{\rm eq}}{\delta\psi_a}
    =
    \int \mathrm{d}x\,
    \sum_{ab}
    \mathcal{J}^{\rm R}_a
    \left(\partial_{\psi_a}\partial_{\psi_b}\mathscr{S}\right)
    \partial_x\psi_b
    \overset{?}{=} \int \mathrm{d}x\, \partial_x \mathscr{G} = 0
\end{equation}
The exact-closure condition asks that this last expression be a total spatial
derivative. We find the condition on $\mathscr{S}$ and $\mathcal{J}^{\rm R}$ that guarantees
this. Define the coefficient $  F_b(\boldsymbol{\psi}) =
\sum_a \mathcal{J}^{\rm R}_a(\boldsymbol{\psi})\, \partial_{\psi_a}
\partial_{\psi_b} \mathscr{S}(\boldsymbol{\psi}).$  Then, for a local $\mathscr{G}(x)$, we obtain, by use
of the chain rule,
\begin{equation}
    \sum_b F_b(\boldsymbol{\psi})\partial_x\psi_b =  \partial_x \mathscr{G}(\boldsymbol{\psi})
    =
    \sum_b
    \frac{\partial \mathscr{G}}{\partial\psi_b}
    \partial_x\psi_b.
\end{equation}
\noindent
Thus it is enough to find $\mathscr{G}$ such that
\begin{equation}
    F_b(\boldsymbol{\psi})
    =
    \frac{\partial \mathscr{G}}{\partial\psi_b}(\boldsymbol{\psi})
    \quad
    \text{for every } b.
\end{equation}
By taking another derivative $\partial_{\psi_c}$ we find a necessary local 
consistency check
\begin{equation}
    \partial_{\psi_c} F_b - \partial_{\psi_b} F_c = 0,
    \quad
    \Rightarrow
    \quad
    \sum_a
    (\partial_{\psi_c} \mathcal{J}^{\rm R}_a ) \,\partial_{\psi_a}\partial_{\psi_b} \mathscr{S}
    -
    \sum_a
    (\partial_{\psi_b} \mathcal{J}^{\rm R}_a)\,\partial_{\psi_a}\partial_{\psi_c} \mathscr{S}
    =
    0,    
\end{equation}
Now identify the Hessian of the local potential  $K_{ab}(\boldsymbol{\psi}) 
= \partial_{\psi_a}\partial_{\psi_b}\mathscr{S}(\boldsymbol{\psi}),$ and the Jacobian
of the reversible current $ A_{ab}(\boldsymbol{\psi}) = 
\partial_{\psi_b}\mathcal{J}^{\rm R}(\boldsymbol{\psi})$.  In this 
parametrization the local consistency reads
\begin{equation}
    \sum_a \left( A_{ac} K_{ab} - A_{ab} K_{ac} \right) = 
    \sum_a \left(  K_{ba} A_{ac} - K_{ca} A_{ab}  \right)
    =0,
    \quad
    \Rightarrow
    \quad
    \boxed{
        KA-(KA)^T = B -B^T = 0.
    }
\end{equation}
\noindent
where in the second line we used the obvious symmetry $K_{ij}=K_{ji}$, and then rewrote the equation in matrix form, and introducing a new matrix $B_{ij} = K_{ia}A_{aj}$. The consistency condition requires
that $B$ must be a symmetric matrix. If this vanishes as an identity in
$\boldsymbol{\psi}$, the integrand in Eq.\,\eqref{eq:deterministic_reversibility}
is a total derivative and stationarity is manifest for periodic fields. This is exactly the Maxwell condition. Under the identification $K=C^{-1}$ we can see this condition is equivalent to
$A(\boldsymbol{\psi})C(\boldsymbol{\psi})=C(\boldsymbol{\psi})A^T(\boldsymbol{\psi})$,
which must be satisfied on the level of every macro state.

This condition can be checked numerically up to a certain order $n$. One 
first imposes the thermodynamic potential, or, equivalently $K_{(n)}$ and the
equation of motion, or, equivalently $A_{(n)}$ up to some order $n$. Then,
the deviation $\Delta=(B)-(B)^T$ can be evaluated on a grid of values. For a two-mode
model this is done on a grid of $\{\psi_1\}$ and $\{\psi_2\}$ and then one may
compute $\max R$ which should be near machine precision if $B$ is symmetric.

\subsection{Automatically stationary models}
The stationarity condition checks when the above integrand can be written
as a total derivative. We can instead suppose an ansatz which automatically
gives stationary EFTs at the bare and dressed level. Introducing new variables 
$\delta \mathcal{S}_{\rm eq}/\delta  \psi_i=\gamma_i(x)$, the stationarity
condition reads
\begin{equation} 
    \sum_a \mathcal{J}_a^{\rm R} \partial_x \gamma_a(x) =\partial_x \mathscr{G}
\end{equation}
The key idea is as follows; since $\boldsymbol{\psi}$ and $\boldsymbol{\gamma}$ 
are conjugate variables to one another, we can equivalently express the current 
$\mathcal{J}_a^{\rm R}[\boldsymbol{\psi}]$ instead as $\mathcal{J}_a^{\rm R}
[\boldsymbol{\gamma}]$. It is then natural to write an expansion in $\gamma_b$
\begin{equation}
    \mathcal{J}_a^{\rm R} = \sum_b \Gamma_{ab} \gamma_b,
\end{equation}
which solves the stationarity if $\Gamma_{ab}=\Gamma_{ba}$. To explicitly 
show this, we compute
\begin{equation}
    \sum_{ab} \Gamma_{ab} \gamma_b \partial_x \gamma_a
    = \frac{1}{2} \sum_{ab}  \left[ \Gamma_{ab} \gamma_b \partial_x \gamma_a + \Gamma_{ba} \gamma_a \partial_x \gamma_b \right]
    = \frac{1}{2} \sum_{ab} \Gamma_{ab} \left[  \gamma_b \partial_x \gamma_a + \gamma_a \partial_x \gamma_b \right]
    = \frac{1}{2} \sum_{ab} \Gamma_{ab} \partial_x \left(\gamma_a \gamma_b \right),
\end{equation}
\noindent
where after the second equality we used $\Gamma_{ab}=\Gamma_{ba}$ and we 
obtain a total derivative. The same derivation works for the general ansatz
\begin{equation}
    \label{eq:master_potential_appendix}
    \mathscr{G}[\mathbf{\gamma}] = \sum_{n\geq 0}\frac{1}{n!}
\sum_{j_{1},\ldots,j_{n}}G_{j_{1},\ldots,j_{n}}\int dx \prod_{i=1}^{n}\gamma_{j_{i}}(x),
\end{equation}
and can be done order by order to show that one obtains a total derivative.

We now construct a minimal model using the linear expansion in $\gamma$ with two modes. We start with a general two-component quartic
potential
\begin{equation}
\begin{split}
    \mathscr{S}(\psi_1,\psi_2)
    =&\frac{1}{2}
    \left(
        K^{(2)}_{11}\psi_1^2
        +2K^{(2)}_{12}\psi_1\psi_2
        +K^{(2)}_{22}\psi_2^2
    \right)+\frac{1}{6}
    \left(
        K^{(3)}_{111}\psi_1^3
        +3K^{(3)}_{112}\psi_1^2\psi_2
        +3K^{(3)}_{122}\psi_1\psi_2^2
        +K^{(3)}_{222}\psi_2^3
    \right)  \\
    &+\frac{1}{24}
    \left(
        K^{(4)}_{1111}\psi_1^4
        +4K^{(4)}_{1112}\psi_1^3\psi_2
        +6K^{(4)}_{1122}\psi_1^2\psi_2^2
        +4K^{(4)}_{1222}\psi_1\psi_2^3
        +K^{(4)}_{2222}\psi_2^4
    \right).
\end{split}
\end{equation}
The conjugate variables are $\gamma_a = \partial \mathscr{S}/\partial \psi_a$. By taking a symmetric $\Gamma = 
\left(\begin{smallmatrix}
    \alpha & \beta \\
    \beta & \delta
\end{smallmatrix}\right)$ 
the reversible current reads
\begin{equation}
    \mathcal{J}^{\rm R}_1=\alpha\gamma_1+\beta\gamma_2,
    \quad
    \mathcal{J}^{\rm R}_2=\beta\gamma_1+\delta\gamma_2.
\end{equation}
Written directly in terms of the physical fields, the completely
general form reads
\begin{equation}
\begin{split}
    \mathcal{J}^{\rm R}_1
    &=
    (\alpha K^{(2)}_{11}+\beta K^{(2)}_{12})\psi_1
    +(\alpha K^{(2)}_{12}+\beta K^{(2)}_{22})\psi_2
    +\frac{1}{2}(\alpha K^{(3)}_{111}+\beta K^{(3)}_{112})\psi_1^2
    \\
    &+(\alpha K^{(3)}_{112}+\beta K^{(3)}_{122})\psi_1\psi_2
    +\frac{1}{2}(\alpha K^{(3)}_{122}+\beta K^{(3)}_{222})\psi_2^2
    +\frac{1}{6}(\alpha K^{(4)}_{1111}+\beta K^{(4)}_{1112})\psi_1^3
    \\
    &+\frac{1}{2}(\alpha K^{(4)}_{1112}+\beta K^{(4)}_{1122})\psi_1^2\psi_2
    +\frac{1}{2}(\alpha K^{(4)}_{1122}+\beta K^{(4)}_{1222})\psi_1\psi_2^2
    +\frac{1}{6}(\alpha K^{(4)}_{1222}+\beta K^{(4)}_{2222})\psi_2^3,
    \\
    \mathcal{J}^{\rm R}_2
    =&
    (\beta K^{(2)}_{11}+\delta K^{(2)}_{12})\psi_1
    +(\beta K^{(2)}_{12}+\delta K^{(2)}_{22})\psi_2
    +\frac{1}{2}(\beta K^{(3)}_{111}+\delta K^{(3)}_{112})\psi_1^2\\
    &
    +(\beta K^{(3)}_{112}+\delta K^{(3)}_{122})\psi_1\psi_2
    +\frac{1}{2}(\beta K^{(3)}_{122}+\delta K^{(3)}_{222})\psi_2^2
    +\frac{1}{6}(\beta K^{(4)}_{1111}+\delta K^{(4)}_{1112})\psi_1^3\\
    &+\frac{1}{2}(\beta K^{(4)}_{1112}+\delta K^{(4)}_{1122})\psi_1^2\psi_2
    +\frac{1}{2}(\beta K^{(4)}_{1122}+\delta K^{(4)}_{1222})\psi_1\psi_2^2
    +\frac{1}{6}(\beta K^{(4)}_{1222}+\delta K^{(4)}_{2222})\psi_2^3.
\end{split}
\end{equation}
\noindent
Note that in this parametrization the (bare) expansion
coefficients in terms of physical fields $\phi_i$ can
be expressed as
\begin{equation}
    A_{i|j} = \Gamma_{ik} K_{kj}^{(2)},
    \quad
    A_{i|jk} = \Gamma_{i \ell} K_{\ell jk}^{(3)},
    \quad
    A_{i|jk\ell} = \Gamma_{i m} K_{m jk\ell}^{(4)}.
\end{equation}
A useful minimal model which exhibits diffusive and superdiffusive transport
can be found within this family. For simplicity, we choose the following
\begin{equation}
    K^{(2)}=\mathbbm{1},
    \qquad
    \Gamma=
    \begin{pmatrix}
        -c & 0\\
        0 & c
    \end{pmatrix},
\end{equation}
\noindent
which gives the linear part of the current $A_{(1)}^\flat = \Gamma$ with
two bare velocities $\pm c$. The smallest cubic part of the potential 
which gives one mode with self-coupling and one mode with no diagonal quadratic
couplings is
\begin{equation}
    K^{(3)}_{111}=-\frac{\lambda}{c},
    \qquad
    K^{(3)}_{122}=\frac{\chi\lambda}{c},
    \qquad
    K^{(3)}_{112}=K^{(3)}_{222}=0,
\end{equation}
with all index permutations understood.  Here $\lambda$ is the self-coupling 
parameter and $\chi$ controls the off-diagonal coupling to the second mode. 
To keep the equilibrium measure confining while adding the fewest
new parameters, take
\begin{equation}
    K^{(4)}_{1111}=K^{(4)}_{2222}=\kappa,
    \qquad
    K^{(4)}_{1112}=K^{(4)}_{1122}=K^{(4)}_{1222}=0,
    \qquad
    \kappa>0.
\end{equation}
Equivalently, the local potential is
\begin{equation}
    \mathscr{S}(\psi_1,\psi_2)
    =
    \frac{1}{2}(\psi_1^2+\psi_2^2)
    -\frac{\lambda}{6c}\psi_1^3
    +\frac{\chi\lambda}{2c}\psi_1\psi_2^2
    +\frac{\kappa}{24}(\psi_1^4+\psi_2^4).
\end{equation}
The reversible current is therefore
\begin{equation}
    \mathcal{J}^{\rm R}_1 =
    -c\psi_1+\frac{\lambda}{2}\psi_1^2-\frac{\chi\lambda}{2}\psi_2^2
    -\frac{c\kappa}{6}\psi_1^3,
    \qquad
    \mathcal{J}^{\rm R}_2=
    c\psi_2+\chi\lambda \psi_1\psi_2 +\frac{c\kappa}{6}\psi_2^3.
\end{equation}
The dissipative current is fixed by the same local potential.  The conjugate
fields are
\begin{equation}
    \gamma_1
    =
    \frac{\partial s}{\partial \psi_1}
    =
    \psi_1-\frac{\lambda}{2c}\psi_1^2
    +\frac{\chi\lambda}{2c}\psi_2^2
    +\frac{\kappa}{6}\psi_1^3,
    \qquad
    \gamma_2
    =
    \frac{\partial s}{\partial \psi_2}
    =
    \psi_2+\frac{\chi\lambda}{c}\psi_1\psi_2
    +\frac{\kappa}{6}\psi_2^3.
\end{equation}
The diagonal deterministic dissipative
part of the current is therefore
\begin{equation}
    \mathcal{J}^{\rm D}_a=-b_a\partial_x\gamma_a.
\end{equation}
Equivalently, written directly in terms of the fields in the total-derivative
form used by the numerical discretization,
\begin{equation}
    \mathcal{J}^{\rm D}_1=
    -b_1\partial_x
    \left[
        \psi_1-\frac{\lambda}{2c}\psi_1^2
        +\frac{\chi\lambda}{2c}\psi_2^2
        +\frac{\kappa}{6}\psi_1^3
    \right],
    \quad
    \mathcal{J}^{\rm D}_2=
    -b_2\partial_x
    \left[
        \psi_2+\frac{\chi\lambda}{c}\psi_1\psi_2
        +\frac{\kappa}{6}\psi_2^3
    \right].
\end{equation}
The stochastic current is added with the matching fluctuation-dissipation
amplitude, $\sigma_{ab} = \delta_{ab} 2b_a$.

\section{Equilibration and stationarity} \label{appendix:equilibration}
We briefly recap the Schr\"odinger versus Heisenberg picture of evolution. In
the former, the evolution is encoded in the measure while the observable $\mathcal{O}$
remains constant. We introduce $\boldsymbol{\psi}=\{\psi_1,\ldots\psi_{N_Q}\}$, and define the expectation values formally as
\begin{equation}
    \langle O \rangle_{\mathcal{P}_t} \equiv 
    \int \mathcal{D}\boldsymbol{\psi}\,O[\boldsymbol{\psi}]\mathcal{P}_t[\boldsymbol{\psi}].
\end{equation}
\noindent
Therefore, at any given time $t$, the observable is averaged over the ensemble $\mathcal{P}_t$.
On the other hand, an alternative perspective, where the evolution is encoded in the observable is as follows. A sample 
$\boldsymbol{\psi}_{0}$ from a distribution $\mathcal{P}_0$ 
is evolved to time $t$ along a noise realization $\boldsymbol{\xi}$. We introduce the shorthand notation  $\boldsymbol{\psi}_t = \boldsymbol{\psi}_t[\boldsymbol{\psi}_0; \boldsymbol{\xi}]$, defining the averaging over noise as $\mathcal{O}[\boldsymbol{\psi}_t]\equiv \int \mathcal{D}\boldsymbol{\xi} \mathcal{P}_{\xi} \mathcal{O}[\boldsymbol{\psi}_t]$ with the measure $\mathcal{P}_{\xi}[\boldsymbol{\xi}]\simeq \exp{(-\tfrac{1}{2}\sum_{a,b}\int \mathrm{d}t\,dx\,\xi_{a}(x,t)[(\sigma^{\flat})^{-1}]_{ab}\xi_{b}(x,t))}$. Averaging over $\mathscr{P}_0$ then defines the dynamical expectation value 
\begin{equation}
    \langle \mathcal{O}(t) \rangle_{\mathcal{P}_{0},\boldsymbol{\xi}}\equiv \int \mathscr{D}\boldsymbol{\psi}_{0}\,\mathcal{P}_{0}[\boldsymbol{\psi}_{0}]\langle \mathcal{O}[\boldsymbol{\psi}_t]\rangle_{\boldsymbol{\xi}}, 
\end{equation}
\noindent
Both give the same expectation values, but it is useful to understand their
differences. We now apply the first picture to probe relaxation to $\mathcal{P}_{\mathrm{eq}}$
and stationarity. The most general measure one can write is
\begin{equation}
    \mathcal{P}_t[\boldsymbol{\psi}] \propto \exp\left[-\mathcal{S}_t[\boldsymbol{\psi}] \right],
    \quad
    \mathcal{S}_t[\boldsymbol{\psi}] = \sum_{n=1}^\infty \frac{1}{n!} \sum_{i_1 \ldots i_n} 
    \int \mathrm{d}x_1\ldots \mathrm{d}x_n  \mathcal{H}_{i_1,\ldots,i_n}(\mathbf{x},t) 
    \psi_{i_1}(x_1) \ldots \psi_{i_n}(x_n),
\end{equation}
where the dynamical vertices $\mathcal{H}_{(n)}\equiv\mathcal{H}_{i_1 \ldots i_n}(\mathbf{x}, t)$
are functions of all $n$ positions and time. The dynamical vertices are Legendre duals 
of equal-time connected correlation functions of the form
\begin{equation}
    \mathscr{C}_{(n)} \equiv
    \mathscr{C}_{i_1\ldots i_n}(\mathbf{x},t)=
    \langle \psi_{i_1}(x_1,t) \ldots \psi_{i_n}(x_n,t) \rangle_{\mathcal{P}_0}^c.
\end{equation}
Note that integrals of $\mathscr{C}_{(n)}$ give the usual definition of thermodynamic
covariances $C_{(n)}$ which are constant in time, while the equal-time functions
$\mathscr{C}_{(n)}$ do have non-trivial time evolution and can be used
to probe convergence to the equilibrium state $\mathcal{P}_{\mathrm{eq}}$.

Before proceeding it is convenient to note that, by imposing translational invariance of
 $\mathcal{H}_{(n)}$
\begin{equation}
    \mathcal{H}_{i_1 \ldots i_n} (x_1 + d, \ldots, x_n + d, t) = 
    \mathcal{H}_{i_1 \ldots i_n} (x_1, \ldots, x_n, t)
    \rightarrow
    \mathcal{H}_{i_1 \ldots i_n} (0, \ldots, r_n, t),
\end{equation}
\noindent
for $d\in\mathbbm{R}$ where we have introduced relative coordinates $r_i=x_i-x$ and $x=x_1$. 
The integrand of the kernel can be rewritten using the new coordinates
\begin{equation}
    \int \mathrm{d}x \mathrm{d}r_2 \ldots \mathrm{d}r_n \,
    \mathcal{H}_{i_1 \ldots i_n} (0, r_2, \ldots, r_n, t)
    \psi_{i_1}(x) \psi_{i_2}(x+r_2) \ldots \psi_{i_n}(x+r_n).
\end{equation}
\noindent
Likewise, the Legendre duals can be expressed as $\mathscr{C}_{i_1 \ldots i_n}(0,r_2,\ldots,r_n)$.

The equilibrium state is reached if $\mathcal{H}_{(n)}$ (or equivalently $\mathscr{C}_{(n)}$) 
become of contact form in the limit $t\to\infty$
\begin{equation}
    \lim_{t\to\infty} \mathcal{H}_{i_1 \ldots i_n}(0,\ldots,r_n,t)=
    K_{i_1\ldots i_n} \prod_{m=2}^n \delta(r_m).
\end{equation}
By inserting this into the definition of the general measure one obtains the
expected equilibrium measure
\begin{equation}
    \begin{split}
    \lim_{t\to\infty} \mathcal{P}_t \sim 
    &\exp\left[
        -\sum_{n=1}^\infty \frac{1}{n!}\sum_{i_1 \ldots i_n} K_{i_1 \ldots i_n}
        \int \mathrm{d}x \mathrm{d}r_2 \ldots \mathrm{d}r_n
         \psi_{i_1}(x) \psi_{i_2}(x+r_2) \ldots \psi_{i_n}(x+r_n) 
         \prod_{m=2}^n \delta(r_m) \right] \\
    &=\exp\left[
        -\sum_{n=1}^\infty \frac{1}{n!}\sum_{i_1 \ldots i_n} K_{i_1 \ldots i_n}
        \int \mathrm{d}x 
        \psi_{i_1}(x) \psi_{i_2}(x) \ldots \psi_{i_n}(x)
         \right]
    = \mathcal{P}_{\mathrm{eq}}.
    \end{split}
\end{equation}
From the point of view of the correlator $\mathscr{C}_{(n)}$, at long times in finite volume $L$,
only contributions at equal positions will contribute 
\begin{equation}
    \mathcal{M}_{(n)}\equiv
    \mathcal{M}_{i_1\ldots i_n}(t)
    =
    \frac{1}{L}\int \mathrm{d}x\,
    \left\langle
    \psi_{i_1}(x,t)\cdots\psi_{i_n}(x,t)
    \right\rangle_{\mathcal{P}_0}^c .
\end{equation}
If the spatial correlator $\mathscr{C}_{(n)}$ has the equilibrium contact form
\begin{equation}
    \mathscr{C}_{i_1\ldots i_n}
    (0,r_2,\ldots,r_n)
    =
    M_{i_1\ldots i_n}
    \prod_{m=2}^n\delta(r_m),
\end{equation}
then $M_{(n)}$ are exactly the Legendre duals of the equilibrium measure vertices
\begin{equation}
    M_{(n)} \equiv M_{i_1 \ldots i_n} = \frac{1}{L} \int \mathrm{d}x 
    \langle \psi_{i_1}(x) \ldots \psi_{i_n}(x)\rangle_{\mathcal{P}_{\rm eq}}^C
\end{equation}
\noindent
Therefore, we can formulate the convergence to the correct equilibrium measure by
\begin{equation*}
    \boxed{
        \begin{split}
            &\mathrm{Condition \ (i): \ Convergence \ of \ } \mathscr{C}_{(n)} \mathrm{\ to \ contact \ form.}   \\
            &\mathrm{Condition \ (ii):} \lim_{t\to\infty} \mathcal{M}_{i_1 \ldots i_n} = M_{i_1 \ldots i_n}.
        \end{split}
    }
\end{equation*}

\newpage

\section{Integration scheme} \label{appendix:numerics}
In this appendix we discuss integration schemes to numerically integrate
stochastic partial differential equations of the form discussed in
the main text
\begin{equation}
    \partial_t \psi_a + \partial_x \mathcal{J}_a[\boldsymbol{\psi}] = 0, 
    \quad 
    \mathcal{J}_a = \mathcal{J}_a^{\rm R} + \mathcal{J}_a^{\rm D} + \xi_a
    \label{eq:continuity_equation}
\end{equation}
\noindent
where $\langle \xi_a(x,t) \rangle_{\boldsymbol{\xi}} = 0$ and $\langle \xi_a(x,t)\xi_b(x',t')\rangle_{\boldsymbol{\xi}} = 2b_a \delta_{ab}
\delta(x-x')\delta(t-t')$, restricting ourselves to diagonal noise. The fields are sampled according to a quartic equilibrium 
distribution
\begin{equation}
    \mathcal{P}_{\rm eq} = Z^{-1} \exp\left[-\mathcal{S}_{\rm eq}\right],
    \quad
    \mathcal{S}_{\rm eq} = \int\mathrm{d}x \mathscr{S}(x),
    \quad 
    \mathscr{S}(x) = \frac{1}{2} \sum_{ij} K_{ij}^\flat \psi_i \psi_j 
    + \frac{1}{6} \sum_{ijk} K_{ijk}^\flat \psi_i \psi_j \psi_k
    + \frac{1}{24} \sum_{ijk\ell} K_{ijk\ell}^\flat \psi_i \psi_j \psi_k \psi_\ell.
\end{equation}
\noindent
Strictly speaking, we must work with a discretized version of $\mathcal{P}_{\rm eq}$, for details see Appendix \ref{appendix:metropolis}.
We will work directly with physical fields $\psi_a(x)$, 
transformations to normal modes $\phi_a(x)$ can be done on the level of individual 
observables, if necessary, as discussed in the main text.

The reversible part of the current $\mathcal{J}_a^{\rm R}$ is given in terms of an
expansion of nonlinearities, while the dissipative part assumes a diagonal gradient form
\begin{equation}
    \mathcal{J}_a^{\rm R} = 
    \sum_{n} \sum_{i_1 \ldots i_n} A^\flat_{a|i_1 \ldots i_n} \psi_{i_1} \ldots \psi_{i_n},
    \quad
    \mathcal{J}_a^{\rm D} =
    - b_a \partial_x \frac{\delta \mathcal{S}_{\rm eq}}{\delta \psi_a}.
\end{equation}
\noindent
We will use the same general approach to integrate Eq.\,\eqref{eq:continuity_equation} 
as Ref.\,\cite{minami2026symmetry} which uses a scheme based on 
\cite{delong2013review,srivastava2023staggered,garcia2024introduction}. The numerical 
computation is done in finite  volume $\Lambda = [-L/2, L/2)$ that is discretized on an equidistant grid $x_j=j \Delta x$, $j=1,\ldots,N$ 
with lattice spacing $\Delta x$. Unless noted otherwise, all simulations use 
$L=N$ ($\Delta x=1$).  The fields are defined on the grid of points $\psi_a(x_j)\equiv
\psi_{a,j}$, while the $\mathcal{J}$ currents  will be defined at interfaces between cells $x_{j\pm 1/2}$. By use of this discretization, the continuity  equation becomes a differential-difference equation
\begin{equation}
    \partial_t \psi_{a,j}(t) = 
    - \frac{\mathcal{J}_{a,j+1/2}(t) - \mathcal{J}_{a,j-1/2}(t)}{\Delta x}
    \equiv \mathrm{J}_{a,j}\left[ \mathcal{J} (t) \right].
\end{equation}
Here $\mathcal{J}_{a,j+1/2}$ is the discrete current, and we additionally introduced the lattice current divergence $\mathrm{J}$ for shorter 
notation, which is simply the nearest-neighbor interface-flux of $\mathcal{J}_{a,j}(t)$.
The discrete current $\mathcal{J}_{a,j}(t)$ is decomposed into a reversible, dissipative and stochastic part
\begin{equation}
    \mathcal{J}_{a,j+1/2}(t) = R_{a,j+1/2}(t) + D_{a,j+1/2}(t) + S_{a,j+1/2}(t),
\end{equation}
each of which is again of flux-preserving form (i.e. defined on the interface) 
and must therefore, for the implementation, be approximated using field values 
on the original grid $\psi_{a,j}(t)$. To simplify the discussion we
join the reversible and dissipative terms into a single, deterministic, term 
$\mathcal{D}_{a,j+1/2}(t) = R_{a,j+1/2}(t) + D_{a,j+1/2}(t)$, which we elaborate 
on later.

The differential-difference system of equations can then be written as
\begin{equation}
    \partial_t \psi_{a,j}( t) = \mathrm{J}_{a,j}\left[\mathcal{D}(t)\right] + \mathrm{J}_{a,j}\left[S(t)\right],
\end{equation}
\noindent
where the stochastic term $\mathrm{J}_{a,j}\left[S(t)\right]$ is defined as a divergence of a stochastic flux
\begin{equation}
    \mathrm{J}_{a,j}[S(t)] = 
    - \frac{\theta_{a,j+1/2}(t) - \theta_{a,j-1/2}(t)}{\Delta x},   
\end{equation}
\noindent
where $\theta_{a,j+1/2}$ are random variables defined on the interfaces.
To integrate the system of equations, we will use the strong  stability-preserving Runge-Kutta scheme of order 3 with a constant
timestep $\Delta t$. In this case, the update is done in three stages. 
The random flux is given as
\begin{equation}
     \theta_{a,j+1/2}(t_n) = 
    \sqrt{\frac{2 b_a}{\Delta x \Delta t}} (\xi_{a,j}^A + \beta_k \xi_{a,j}^B),
\end{equation}
where $\xi_{a,j}^A,\xi_{a,j}^B \sim \mathcal{N}(0,1)$ are Gaussian variables with unit
variance, and each of the coefficients $\beta_k$ are used at the appropriate (1st, 2nd, or 3rd) stage. They read
\begin{equation}
    \beta_1 = \frac{2 \sqrt{2} + \sqrt{3}}{5}, 
    \quad
    \beta_2 = \frac{-4 \sqrt{2} + 3 \sqrt{3}}{5},
    \quad
    \beta_3 = \frac{\sqrt{2} - 2\sqrt{3}}{10}.
\end{equation}
The update from $t_n$ to $t_{n+1}$ is a standard Runge-Kutta procedure 
with an added stochastic term. The three stages read
\begin{equation}
    \begin{split}
        \psi_{a,j}( t_{n+1/3}) &= 
        \psi_{a,j}( t_n) 
        + \Delta t \mathrm{J}_{a,j}[\mathcal{D}(t_n)] 
        + \Delta t \mathrm{J}_{a,j}[S(t_n)],
        \\
        \psi_{a,j}( t_{n+2/3}) &=
        \frac{3}{4} \psi_{a,j}( t_n) 
        + \frac{1}{4} \left[\psi_{a,j}( t_{n+1/3})
        + \Delta t \mathrm{J}_{a,j}\left[\mathcal{D}(t_{n+1/3})\right]
        + \Delta t \mathrm{J}_{a,j} \left[S(t_{n+1/3})\right]
        \right],
        \\
        \psi_{a,j}( t_{n+1}) &=
        \frac{1}{3} \psi_{a,j}( t_n) 
        + \frac{2}{3} \left[\psi_{a,j}( t_{n+2/3})
        + \Delta t \mathrm{J}_{a,j} \left[\mathcal{D}(t_{n+2/3})\right]
        + \Delta t \mathrm{J}_{a,j} \left[S(t_{n+2/3})\right]
        \right].
        \\
    \end{split}
\end{equation}

\subsection{Cyclic case}
The main problem of integration is the construction of a spatial
discretization scheme. There exist many discretizations which, in
the continuum limit $\Delta x \to 0$, have the same properties as
the parent field theory. There are exceptional discretizations, that can preserve the same properties for any $\Delta x$. We first discuss 
the simplest case, where 
\begin{equation}
      \mathcal{J}_a^{\rm R}[\boldsymbol{\psi}] = 
      \sum_{i_1} A^\flat_{a|i_1} \psi_{i_1} + 
      \sum_{i_1, i_2} A^\flat_{a | i_1 i_2} \psi_{i_1} \psi_{i_2},
\end{equation}
\noindent
with the additional property that $A_{a | i_1 i_2}^\flat$ is cyclic in all three 
indices. As discussed in the main text, the stationary measure
$\mathcal{P}_{\mathrm{eq}}$ of the given SPDE with the given constraint is purely Gaussian. This equation of motion admits a convenient 
discretization of the deterministic part \cite{roy2024universality,minami2026symmetry} which, in the diagonal case $A_{a|i_1}^\flat = \delta_{a,i_1} c_a$ and $K_{ab}^\flat = \delta_{ab}$, reads as follows
\begin{equation}
     \mathcal{D}_{a,j+1/2} = 
     c_a \frac{\psi_{a,j} + \psi_{a,j+1}}{2}
     + \sum_{bc} A_{a|bc}^\flat\left[
       \frac{\psi_{b,j} \psi_{c,j+1} + \psi_{b,j+1} \psi_{c,j}}{12}+
       \frac{\psi_{b,j} \psi_{c,j}   + \psi_{b,j+1}\psi_{c,j+1} } {6} 
	     \right]
	     - b_a \frac{\psi_{a,j+1} - \psi_{a,j}}{\Delta x},
\end{equation}
This discretization exactly preserves the (lattice) counterpart of $\mathcal{P}_{\rm eq}$ \cite{roy2024universality}, 
and can thus be understood as an exact differential-difference version of Eq. 
\,\eqref{eq:continuity_equation}.

As discussed in Appendix \ref{appendix:equilibration}, we can numerically
test stationarity of the integrator by computing correlators of type 
$\mathscr{C}_{(n)}$ and $\mathcal{M}_{(n)}$. If the initial conditions are 
sampled according to $\mathcal{P}_{\rm eq}$ the correlators $\mathscr{C}_{(n)}$ must
remain of contact form, and $\mathcal{M}_{(n)}$ must retain their equilibrium
values at all times. To showcase the test, we consider a two-mode model with
$c_a = \pm 0.1$ and $\lambda_i \in \mathbbm{R}$ with the Hessian
\begin{equation}
    A_{1|bc}^\flat = 
    \begin{pmatrix}
        0 & \lambda_2 \\
        \lambda_2 & \lambda_3
    \end{pmatrix},
    \quad
    A_{2|bc}^\flat = 
    \begin{pmatrix}
        \lambda_2 & \lambda_3 \\
        \lambda_3 & \lambda_4
    \end{pmatrix},
\end{equation}
\noindent
In this case, $\mathscr{C}_{(n>2)}
= 0$, and $M_{(n>2)} = 0$ which gives a useful test case. Furthermore $\mathscr{C}_{(2)}
= C_{ab} \delta(x-y)$ and $C_{ab} = \delta_{ab}$. For the simulation we consider
the values of parameters $\lambda_2 = 0.01$, $\lambda_3 = 1.0$ and $\lambda_4 = 2.0$.

\begin{figure}[h!]
    \centering
    \includegraphics[width=1.0\textwidth]{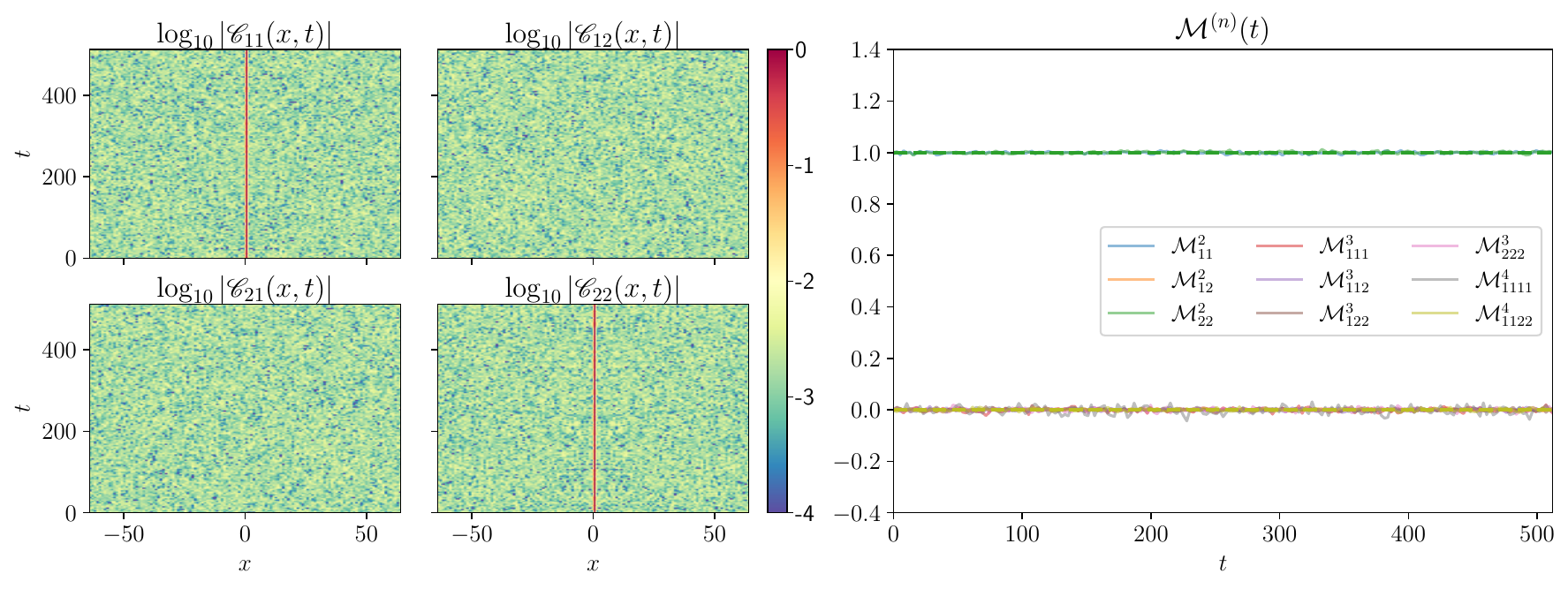}
    \caption{Stationarity of a Gaussian equilibrium state $\mathcal{P}_{\mathrm{eq}}$ under 
    dynamics in the cyclic test model. The left panel shows evolution of $\mathscr{C}^{(2)}$ 
    which preserves contact form, while the right panel shows the dynamics of 
    equal-time correlators $\mathcal{M}^{(n)}(t)$ which evidently fluctuate around 
    expected equilibrium values that are denoted by dashed lines (determined analytically).
    Simulation parameters: $\Delta t = 0.04$, number of samples $1000$, $N=128$.}
    \label{fig:stationarity_cyclic}
\end{figure}

\newpage
\subsection{General case}
\begin{figure}[h!]
    \centering
    \includegraphics[width=0.5\textwidth]{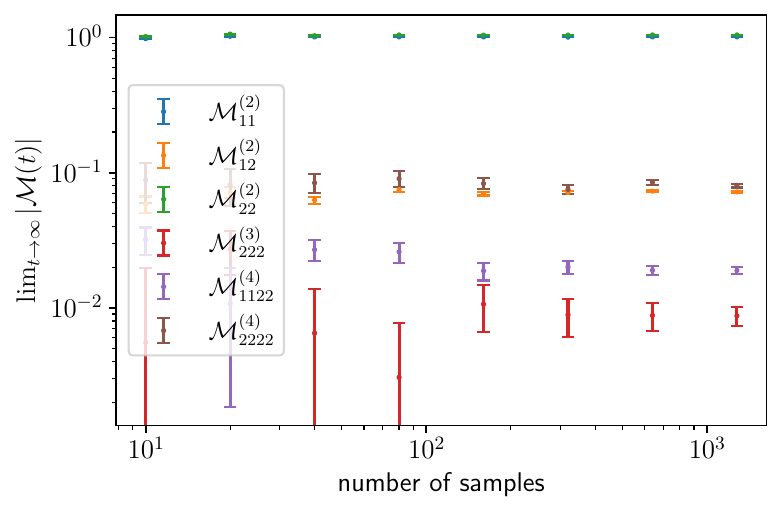}
    \caption{Dynamical generation of nonzero cumulants $\mathcal{M}_{(n>2)}$ under
    dynamics with a non-cyclic Hessian. The equal-time correlators $\mathcal{M}_{(n)}(t)$ are
    evaluated in equilibrium with an increasing number of samples. The Gaussian prediction 
    $\lim_{t\to\infty}\mathcal{M}_{ab} = \delta_{ab}$ and $\lim_{t\to\infty}\mathcal{M}_{(n>2)} = 0$
    are broken. Simulation parameters: $\Delta t = 0.04$, $N=128$.}
    \label{fig:nonstationarity_cyclic}
\end{figure}

In the general case, where $A_{(2)}^\flat$ is not cyclic, the measure does not truncate
at the Gaussian level. Numerically, this can be shown by taking a random non-cyclic Hessian,
e.g.
\begin{equation}
    A_{1|bc}^\flat = 
    \begin{pmatrix}
        0.867347 & -0.901744 \\
        -0.901744 & -0.494479
    \end{pmatrix},
    \quad
    A_{2|bc}^\flat = 
    \begin{pmatrix}
        -0.902914 & 0.864401 \\
        0.864401  & 2.21188
    \end{pmatrix},
\end{equation}
and computing $M_{(n)}$ which start to deviate from the prediction $M_{ab} = \delta_{ab}$
and $M_{(n>2)} = 0$. Note that $A_{(2)}^\flat$ must still be symmetric in the lower indices.
The equation of motion is integrated with the cyclic integrator described in the previous section, until the cumulants converge in time ($t=128$)
$\lim_{t\to\infty} \mathcal{M}_{(n)}(t)$, dynamically generating a non-Gaussian state, which is then sampled with an increasing number of samples to determine the equilibrium values of the cumulants.
As pictured in Fig.\,\ref{fig:nonstationarity_cyclic}, a number of cumulants become non-zero, confirming the deviation from Gaussianity.

The new terms that appear in the stationary measure must be included in the equation of motion to respect the FDR and stationarity.
As discussed in the main text, we truncate the measure at $K_{(4)}^\flat$, 
which in general allows for terms of the form $A_{(3)}^\flat$ in the equation 
of motion to satisfy stationarity. Note that, even in the case $A_{(3)}^\flat=0$, 
the gradient form of the dissipative current $ \mathcal{J}_a^{\rm D} = - b_a \partial_x 
\frac{\delta \mathcal{S}_{\rm eq}}{\delta \psi_a}$ requires the addition of new 
dissipative terms which we discretize in the following way. Define the
thermodynamic force on the lattice
\begin{equation}
    \gamma_{a,j} = \frac{\delta \mathcal{S}_{\rm eq}}{\delta \psi_a}
    = \sum_b K_{ab}^\flat \psi_{b,j}
    +\frac{1}{2}\sum_{b,c}K_{abc}^\flat \psi_{b,j}\psi_{c,j}
    +\frac{1}{3!}\sum_{b,c,d}K_{abcd}^\flat
    \psi_{b,j}\psi_{c,j}\psi_{d,j}.
\end{equation}
The nonlinear dissipative interface flux is then discretized as
\begin{equation}
    D^{a}_{j+1/2}
    =
    -b_a\,\frac{\gamma_{a,j+1}-\gamma_{a,j}}{\Delta x},
    \qquad
    \mathcal{J}_{a,j}[D]
    =
    -\frac{D_{a,j+1/2}-D_{a,j-1/2}}{\Delta x}.
 \end{equation}
\noindent
In the numerical scheme these terms act to stabilize the evolution, as
nonlinearities are suppressed. The general form of the dissipative interface flux 
is the nearest-neighbour difference of the form $\mathrm{J}_{a,j}\left[D(t)\right]$ as discussed 
above.

\begin{figure}[h!]
    \centering
    \includegraphics[width=0.49\textwidth]{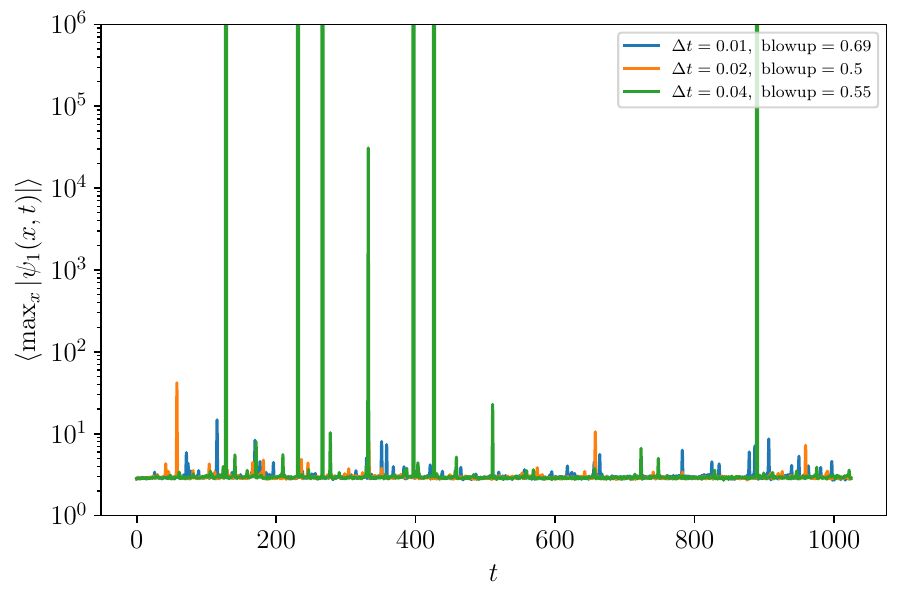}
    \includegraphics[width=0.49\textwidth]{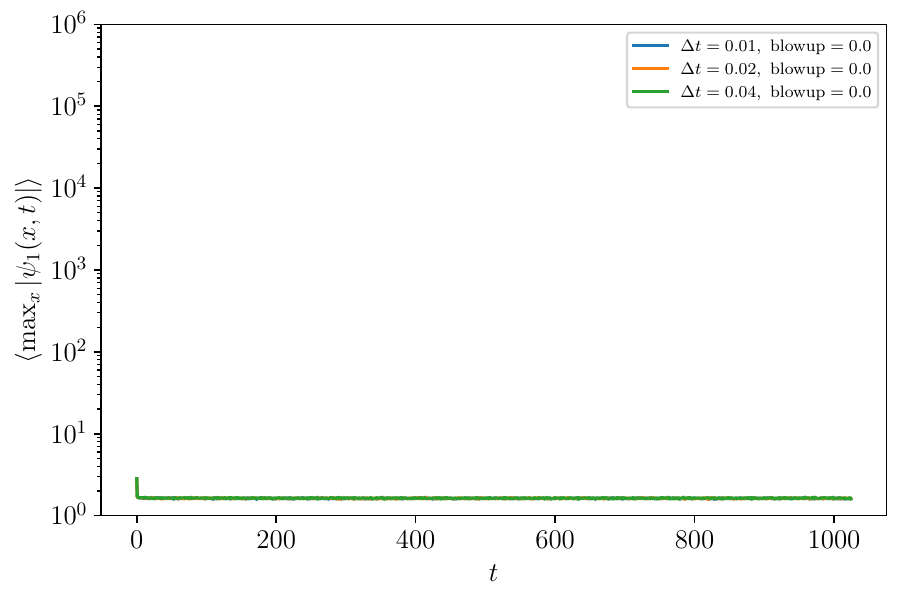}
    \caption{Stability of the cyclic (left panel) and the consistent (right panel) numerical schemes.   The panels show $\max_{x}\psi_1(x,t)$ averaged over 100 initial conditions as a function of time for increasing $\Delta t$. The legend denotes the used timestep $\Delta t$ and the fraction of trajectories that exhibited instability $\max_{x}\psi_1(x,t)>10^6$. The consistent scheme stabilizes the dynamics to a large degree.  Parameters: $N=128, c=0.3,\kappa =2$, number of samples $100$.}
    \label{fig:stability}
\end{figure}

A second point to be made is that strongly non-cyclic Hessians can, by use of the Gaussian preserving scheme, become unstable easily. Take the following Hessians
\begin{equation}
    A_{1|bc}^\flat = 
    \begin{pmatrix}
        2g & 0 \\
        0 & -g
    \end{pmatrix},
    \quad
    A_{2|bc}^\flat = 
    \begin{pmatrix}
        2g & 0 \\
        0  & g
    \end{pmatrix},
\end{equation}
as well as add an advective term $A_{i|j}^\flat = \mathrm{diag}(\pm c)$ for each mode. In contrast to the case where the measure is Gaussian where the integrator is stable even up to $\Delta t$ up to $0.1$ or more (details depending on the parameters), this case can exhibit instability easier for a much wider range of parameters. As can be seen in Fig.\,\ref{fig:stability}, the inclusion of quartic terms in $\mathcal{S}_{\rm eq}$ and, consequently, the nonlinear diffusive terms, acts to stabilize the dynamics (of what we call the \emph{consistent} scheme) by introducing nonlinear damping.

Furthermore, in the general case $A_{(3)}^\flat \neq 0$ so we must prescribe a
discretization procedure for the new term, which is not fixed strictly
by thermodynamic relations. We will now directly specialize the new discretization to the minimal model considered in the main text (c.f the section \textit{Minimal models of nonlinear EFTs}). The discretization of the reversible current is done
order by order $R_{a,j+1/2} = \sum_{i=1}^3 R_{a,j+1/2}^{(i)}$ and makes use of centered endpoint averages
\begin{equation}    
    \begin{split}
    R_{a,j+1/2}^{(1)}  &= c_a \frac{\psi_{a,j}+\psi_{a,j+1}}{2},
    \quad
    R_{a,j+1/2}^{(2)} = \sum_{b,c}A^\flat_{a|bc}\,
    \frac{\psi_{b,j}\psi_{c,j}+\psi_{b,j+1}\psi_{c,j+1}}{4},\\
    &R_{a,j+1/2}^{(3)}
    =
    \frac{1}{3!}\sum_{b,c,d}A_{a|bcd}^\flat\,
    \frac{
    \psi_{b,j}\psi_{c,j}\psi_{d,j}
    +
    \psi_{b,j+1}\psi_{c,j+1}\psi_{d,j+1}
    }{2}.
    \end{split}
\end{equation}
Note that we also used a different discretization of the quadratic term than previously. This gives the correct continuum current and is cheap to evaluate, 
although by itself it does not guarantee that the finite lattice admits 
exactly the same lattice equivalent of the continuum equilibrium measure at finite $\Delta x$.

\begin{figure}[h!]
    \centering
    \includegraphics[width=1.0\textwidth]{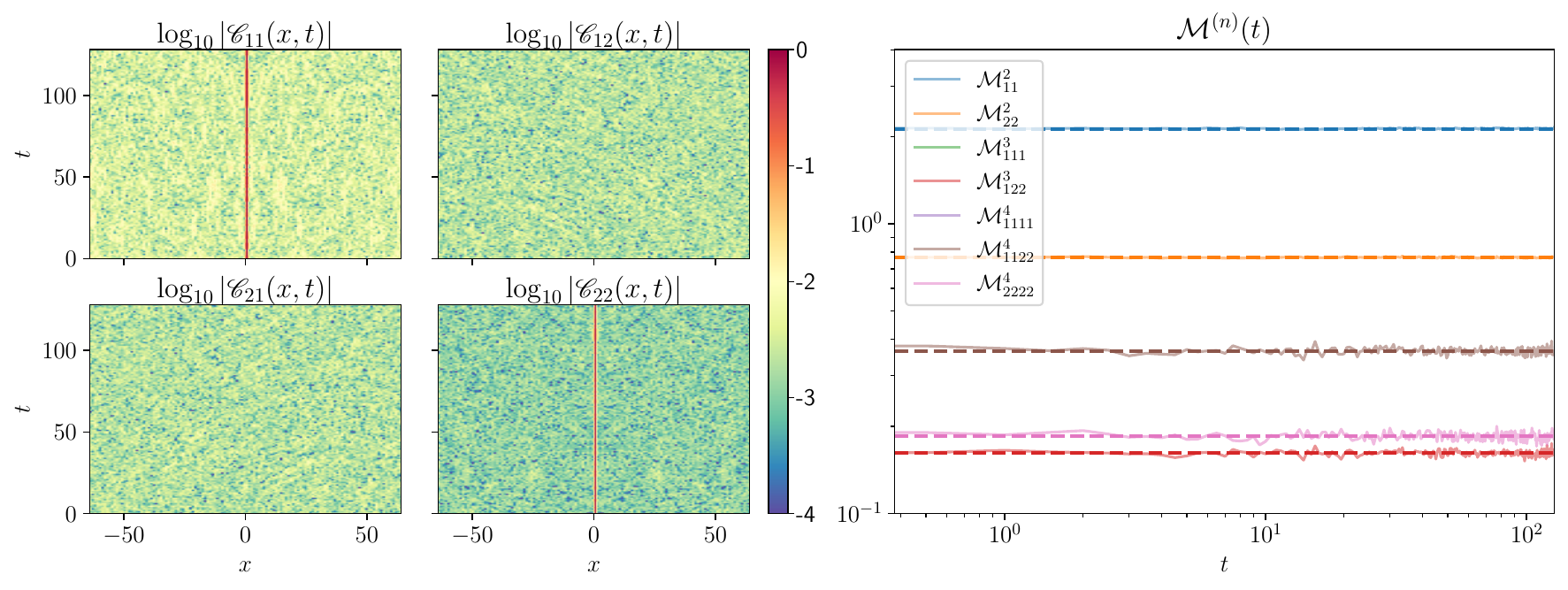}
    \caption{Stationarity of $\mathcal{P}_{\mathrm{eq}}$ under dynamics in the minimal
    model. The left panel shows evolution of $\mathscr{C}^{(2)}$ which preserves
    contact form, while the right panel shows the dynamics of non-zero equal-time correlators
    $\mathcal{M}^{(n)}(t)$ which evidently fluctuate around equilibrium values
    that are denoted by dashed lines (determined by quadrature).
    Simulation parameters: The parameters  $\mathbf{h}=(-1.1303, 0)$  are chosen such that 
    $\langle \psi_i\rangle_h = 0$ (see Appendix \ref{appendix:metropolis}), 
    $\Delta t = 0.01$, number of samples $1000$, $N=128$.}
    \label{fig:stationarity}
\end{figure}

One can again test that the correlators of type $\mathscr{C}_{(n)}$ remain
of contact form, and that strict equilibrium values of $\lim_{t\to\infty}
\mathcal{M}_{(n)}(t)= M_{(n)}$ are matched. This is done by sampling states 
from the equilibrium distribution at zero-mean using a Metropolis procedure 
(as described in Appendix 
\ref{appendix:metropolis}), evolving them with the integrator described above,
and numerically checking the listed requirements. An example for the minimal model 
is shown in Fig.\,\ref{fig:stationarity}, which shows that $\mathscr{C}^{(2)}$ 
remains of contact form (we do not explicitly show higher $n$), and
$\mathcal{M}^{(n)}(t)$ up to $n=4$ do not deviate from equilibrium values denoted by dashed lines (which are determined by quadrature of the measure $\mathcal{S}_{\rm eq}$, more details in Appendix 
\ref{appendix:metropolis}),
confirming stationarity on the observed timescale. We show only non-zero  $\mathcal{M}_{(n)}$ which are unique ($ \mathcal{M}_{212}=\mathcal{M}_{221}=\mathcal{M}_{122}$ and $\mathcal{M} _{1212}=\mathcal{M}_{1122}=\mathcal{M}_{1221}=\mathcal{M}_{2112}=\mathcal{M}_{2121}=\mathcal{M}_{2211}$). Qualitatively this is evidence that the stationary state is stationary and stable under the discretized dynamics, though for a strict proof the equivalence of all $\mathscr{C}^{(n)}$ and $\mathcal{M}^{(n)}$ 
must be shown.

\textbf{Dynamical structure factor.} The dynamical structure factor is computed on a grid $x=j\Delta x$ and $t=m\Delta t$ with $N$ points in space and $N_t$ points in time. Due to stationarity of the state, we can additionally average over  all shifts in space  and time
\begin{equation}
    S_a(j\Delta x,m\Delta t) =\frac{1}{N(N_t-m)}  \sum_{r=1}^N\sum_{k=1}^{N_t-m} \left\langle \psi_{a,j+r}( (m+k)\Delta t) \psi_{a,r}(k\Delta t)\right\rangle_{\mathcal{P}_{\rm eq}, \boldsymbol{\xi}}.
\end{equation}
\noindent
This is efficiently implemented via a Fast Fourier transform. The spatial component is periodic ($j+r$ understood modulo $N$), and therefore warrants the use of a circular transform, while the time component is a linear transform (the data is padded to twice its size with zeros). Strictly speaking, the averaging is done in the discretized tilted ensemble $\mathcal{P}_h$, as defined in Appendix \ref{appendix:metropolis}, where $\mathbf{h}$ is chosen such that $\langle \psi_{a,j}\rangle=0$.

\section{Metropolis} \label{appendix:metropolis}
This section will be dedicated to discussion of a Metropolis sampling procedure. 
The continuum equilibrium measure for $\boldsymbol{\psi}(x)=\{\psi_1(x),\ldots,\psi_{N_Q}(x)\}$ 
discussed  in the main text is
\begin{equation}
    \mathcal{P}_{\mathrm{eq}} = Z^{-1} \exp(- \mathcal{S}_{\rm eq}[\boldsymbol{\psi}] ), 
    \quad
    \mathcal{S}_{\rm eq} [\boldsymbol{\psi}]= \int \mathrm{d}x\, \mathscr{S}[\boldsymbol{\psi}],
    \quad 
    \mathscr{S}[\boldsymbol{\psi}] = \sum_{n=1}^\infty \frac{1}{n!} 
    \sum_{i_1,\ldots,i_n} K_{i_1\ldots i_n}^\flat \psi_{i_1}(x)\ldots \psi_{i_n}(x),
    \label{eq:continuum_equilibrium_distribution}
\end{equation}
\noindent
where $Z$ is the normalization. The continuum equilibrium measure can be discretized on a lattice with $N$ points 
and lattice spacing $\Delta x$ as a distribution of fields $\psi_{a,j}=\psi_a(j \Delta x)$ at discrete lattice points $j$
\begin{equation}
    \mathcal{P}_N = Z^{-1} \exp(-\mathcal{S}_N),
    \quad 
    \mathcal{S}_N = \Delta x \sum_{j=1}^N \mathscr{S}_j,
    \quad
    \mathscr{S}_j = \mathscr{S}(j\Delta x),
    \label{eq:lattice_equilibrium_distribution}
\end{equation}
which can be factorized into independent one-site contributions 
to the lattice equilibrium measure $\mathcal{P}_N$
\begin{equation}
    \mathcal{P}_N = Z^{-1} \exp\left[-\Delta x \sum_j \mathscr{S}_j\right] = 
    Z^{-1} \prod_{j=1}^N \exp\left[- \Delta x \mathscr{S}_j\right] = 
    Z^{-1} \prod_{j=1}^N \mathcal{P}_j, \quad \mathcal{P}_j\equiv \exp[-\Delta x \mathscr{S}_j],
\end{equation}
\noindent
In the special case when the  equilibrium distribution $\mathcal{P}_j$ is Gaussian, it can be 
sampled by drawing samples from a Gaussian distribution with covariance $C^{(2)}/a$.
For a general distribution, one can still perform efficient sampling, as the one-site
measure $\mathcal{P}_j$ is ultra-local, by using a Metropolis algorithm. We begin with a 
Gaussian random  variable $\boldsymbol{\psi}_j$ and propose a random update vector 
$\mathbf{D}_j=\{d_{aj}\}_{a=1}^{N_Q}$ with $d_{aj} \sim \mathcal{N}(0,1)$
\begin{equation}
    \boldsymbol{\psi}'_j = \boldsymbol{\psi}_j + \sigma \mathbf{D}_j,
\end{equation}
\noindent
which are accepted with a probability
\begin{equation}
    p_{\mathrm{acc}} = 
    \min(1, \exp\left[- \Delta x \mathscr{S}_j \left(\boldsymbol{\psi}_j'\right) 
            + \Delta x \mathscr{S}_j\left(\boldsymbol{\psi}_j\right)]\right).
\end{equation}
\noindent
This is efficient due to the complete locality of $\mathscr{S}$ with $\sigma$ acting only 
as an efficiency parameter. After a short initial burn-in period, this Markov chain
samples from the distribution $\mathcal{P}_j$. In the simulation, we initialize a chain length $N$ of Gaussian random variables with variance 1. We then perform an update on each site of the chain, which constitutes a single sweep. To sample an initial configuration we perform 2000 sweeps.

\textbf{Zero-mean tilt convention} While the constraint 
$\int\mathrm{d}x\,\psi_a=0$ is equivalent to $\langle \psi_a \rangle=0$ 
in the thermodynamic limit (for a homogeneous model), it is not in finite volume $L$. If the constraint is satisfied, the finite volume system exhibits fluctuations of the charge on the order of 
$\mathcal{O}(L^{-1/2})$ as the total charge will fluctuate as $\mathcal{O}(L^{1/2})$.

\begin{figure}[h!]
    \centering
\includegraphics[width=1.0\textwidth]{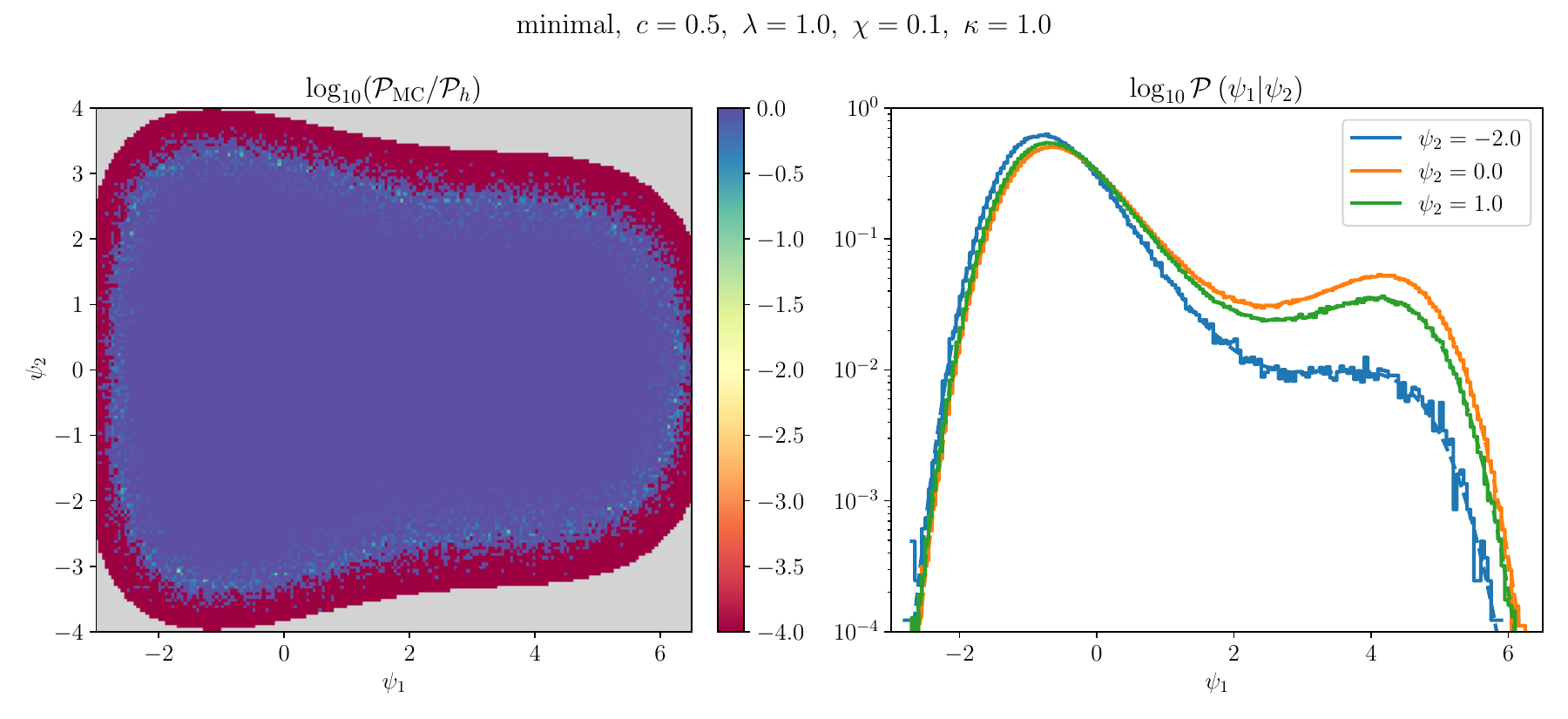}
    \caption{The left panel shows a ratio of the exact and Metropolis-sampled 
    distributions of $\psi$ for the minimal model ($c=0.5, \lambda=1.0,
    \kappa=1.0, \chi=0.1$).
    The grey area shows the region where $\mathcal{P}_h<10^{-8}$. The right panel 
    shows conditional probabilities (full lines) at given $\psi_2$  (a finite bin 
    width 0.05 is taken in the $\psi_2$ direction) compared to exact predictions 
    (dashed lines). The parameters  $\mathbf{h}=(-1.1303, 0)$  are chosen such that 
    $\langle \psi_a\rangle_h = 0$ (see text above). Number of samples is $10^8$.}
    \label{fig:metropolis_distributions}
\end{figure}

To this end, we introduce Legendre parameters $h_a$ (chemical potentials) that can be optimized such that the constraint $\langle \psi_a \rangle=0$ is satisfied. 
We call this the \emph{tilted} ensemble, which reads
\begin{equation}
    \mathcal{P}_h = \prod_{j=1}^N Z_h^{-1} \exp\left(-\Delta x \mathscr{S}_j + \Delta x\sum_a h_a \psi_{a,j} \right),
\end{equation}
\noindent
The tilt provides a constant spatially constant force, whose derivative vanishes, preserving stationarity. To determine the tilt we solve the following one-site problem
\begin{equation}
    \langle \psi_a \rangle_{\mathcal{P}_h} = m_a,
    \label{eq:tilt-root-equation}
\end{equation}
where in the simulations below we choose the target mean $m_a=0$.  The averages
in Eq.\,\eqref{eq:tilt-root-equation} are local equilibrium averages with
respect to $\mathcal{P}_h$. This can be solved iteratively, for example with 
Newton iteration. In this case, it is particularly simple because the 
derivative of the mean with respect to the tilt is the connected covariance 
matrix,
\begin{equation}
    \frac{\partial}{\partial h_b}
    \langle \psi_a\rangle_{\mathcal{P}_h}
    =
    \langle \psi_a\psi_b\rangle_{\mathcal{P}_h}
    -
    \langle \psi_a\rangle_{\mathcal{P}_h}
    \langle \psi_b\rangle_{\mathcal{P}_h}
    \equiv \Delta x\, C_{ab}(\mathbf{h}).
\end{equation}
Therefore, starting from an initial guess $\mathbf{h}^{(0)}$,
 and a target $\mathbf{m}=(m_a)_{a=1}^{N_Q}$, one updates
\begin{equation}
    \mathbf{h}^{(n+1)}
    =
    \mathbf{h}^{(n)}
    +
\left[aC(\mathbf{h}^{(n)})\right]^{-1}
    \left[
        \mathbf{m}
        -
        \langle \mathbf{\Psi}_j\rangle_{\mathcal{P}_{\mathbf{h}^{(n)}}}
    \right].
    \label{eq:tilt-newton-update}
\end{equation}
For the two-mode models the required one-site averages $\langle \psi_a
\rangle_{\mathcal{P}_h}$ and $\langle \psi_a\psi_b\rangle_{\mathcal{P}_h}$
(equivalently, the connected covariance $C_{ab}(\mathbf h)$)
are evaluated by quadrature over $(\psi_1,\psi_2)$, which makes the iteration 
noise-free. The Metropolis algorithm is then used only after the tilt has 
been fixed, to draw independent lattice initial conditions from the 
tilted product measure. A distribution of samples drawn using the Metropolis algorithm is compared with the exact tilted ensemble of the minimal model is shown in Fig.~\ref{fig:metropolis_distributions} where matching is observed.

\textbf{Numerical quadrature} For the minimal models considered, due to the 
exact closure of the two-dimensional measure $\mathcal{P}_{\mathrm{eq}}$ we can 
efficiently numerically compute expectation  values of observables $O(\psi_1, 
\psi_2)$ via quadrature
\begin{equation}
    \langle O \rangle_{\mathcal{P}_{\mathrm{eq}}} = 
    Z^{-1} \iint \mathrm{d}\psi_1 \mathrm{d}\psi_2 \, O(\psi_1,\psi_2) 
    \mathcal{P}_{\mathrm{eq}}(\psi_1,\psi_2), 
    \quad Z = \,
    \iint \mathrm{d}\psi_1 \mathrm{d}\psi_2 \mathcal{P}_{\mathrm{eq}}(\psi_1,\psi_2).
\end{equation}
The integral has to be restricted to a finite region $\psi_i \in (-\psi_{\mathrm{max}},
\psi_{\mathrm{max}})$ for which we choose $\psi_{\mathrm{max}}=10$, and we discretize
$\psi_i$ on this region using an equidistant grid of $1001$ points unless stated
otherwise. Using these parameters, the obtained static two- and three-point covariances have converged to an absolute precision of $<10^{-12}$ (i.e. if the number of grid points is doubled, the results change by $<10^{-12}$).

\end{document}